\documentclass[aps,pra,twocolumn,a4paper,showpacs]{revtex4}
\usepackage{amsmath}
\usepackage{amssymb}

\newcommand{\beq}{\begin{equation}}
\newcommand{\eeq}{\end{equation}}
\newcommand{\cond}{\Phi_0 (\mathbf{r})}
\newcommand{\ccond}{\Phi_0^* (\mathbf{r})}
\newcommand{\gcond}{\Phi_2 (\mathbf{r})}
\newcommand{\gccond}{\Phi_2^* (\mathbf{r})}
\newcommand{\cL}{\mathcal{L}}

\newcommand{\hLGP}{\hat{L}_{0}^{\text{(GP)}}(\mathbf{r})}
\newcommand{\MGP}{M_{0}^{\text{(GP)}}(\mathbf{r})}

\newcommand{\cLGP}{\mathcal{L}^{\text{(GP)}}}
\newcommand{\hsp}{\hat{H}_{\text{sp}}}
\newcommand{\nt}{\tilde{n}}
\newcommand{\mt}{\tilde{m}}
\newcommand{\bfr}{\mathbf{r}}
\newcommand{\bfrp}{\mathbf{r'}}
\newcommand{\rt}{(\mathbf{r},t)}
\newcommand{\rpt}{(\mathbf{r'},t)}
\newcommand{\rr}{(\mathbf{r},\mathbf{r'})}
\newcommand{\rrt}{(\mathbf{r},\mathbf{r'},t)}
\newcommand{\rrw}{(\mathbf{r},\mathbf{r'},\omega)}

\newcommand{\rw}{(\mathbf{r},\omega)}
\newcommand{\rmw}{(\mathbf{r},-\omega)}

\newcommand{\dl}{\delta \lambda}
\newcommand{\intdr}{\int \!\! d \mathbf{r} \,}
\newcommand{\intdrp}{\int \!\! d \mathbf{r'} \,}
\newcommand{\bpm}{\begin{pmatrix}}
\newcommand{\epm}{\end{pmatrix}}

\begin{document}

\title{The response of Bose-Einstein condensates to external perturbations at finite temperature}
\author{S. A. Morgan}
\email[]{sam@theory.phys.ucl.ac.uk}
\affiliation{Department of Physics and Astronomy, University College London, Gower Street, London WC1E 6BT, UK}

\date{\today}

\begin{abstract}

We present a theory of the linear response of a Bose-condensed gas to external perturbations at finite temperature. The theory developed here is the basis of a recent quantitative explanation of the measurements of condensate excitations and decay rates made at JILA [D. S. Jin \textit{et al.}, Phys. Rev. Lett. {\bf 78}, 764 (1997)]. The formalism is based on a dynamic, number-conserving, mean-field scheme and is valid in the collisionless limit of well-defined quasiparticles.
The theory is gapless, consistent with the generalized Kohn theorem for the dipole modes, and includes the time-dependent normal and anomalous averages, Beliaev and Landau processes, and all relevant finite size effects. The important physical process where the thermal cloud is driven directly by the external perturbation is explicitly included. This is required for consistency with the dipole modes and is also needed to explain the JILA observations. 

\end{abstract}


\maketitle


\section{Introduction} \label{INTRODUCTION}

Some of the earliest experiments on Bose-Einstein condensates (BEC) in trapped gases involved measuring the energies and decay rates of condensate excitations \cite{Jin96,Jin97,Mewes96b,StamperKurn98}. These measurements provide a sensitive probe of condensate dynamics and a unique opportunity for quantitative tests of thermal field theories. For this reason condensate excitations continue to attract widespread interest, and a variety of new measurements have been made, notably for scissors mode oscillations \cite{Marago01}, in highly elongated traps \cite{Chevy02}, via Bragg spectroscopy \cite{Steinhauer03}, and in vortex condensates \cite{Bretin03}. In this paper we derive a second order quantum field theory which can be applied to these experiments. This theory has recently been used to obtain good agreement with finite temperature measurements made at JILA in 1997 \cite{Jin97}, which have been the subject of much discussion in the field \cite{Morgan03a}.

Theoretical calculations of condensate excitations in dilute Bose gases are often based on the quasiparticle description introduced by Bogoliubov \cite{Bogoliubov47}. However, 
accurate calculations beyond the Bogoliubov theory are difficult for a variety of reasons. First, the various higher order effects which occur have to be treated consistently to avoid problems of infra-red divergences and to obtain a gapless spectrum, as required by the Hugenholtz-Pines theorem \cite{Hugenholtz59}. Second, one must in principle deal simultaneously with the strong modification of low-energy states due to interactions and the substantial single-particle effects that are present at finite temperature. Third, a dynamic description of both condensed and non-condensed atoms and their mutual interaction is required. While the condensate dynamics is well described by a single nonlinear equation (the Gross-Pitaevskii equation or GPE \cite{GPE}), evolving the non-condensate is a much more complicated numerical problem. Finally, experiments in trapped gases may have substantial finite size effects so that approximations valid in the thermodynamic limit may not be appropriate.

The theory we develop in this paper addresses all these issues using a full quasiparticle description of the non-condensate. The dynamic coupling between condensed and uncondensed atoms is described using perturbation theory. It has been stated in the literature that a perturbative calculation can not explain the JILA measurements \cite{Reidl00,Jackson02}, but our recent calculations show that this is not the case \cite{Morgan03a,Morgan03c}. A perturbative approach is perfectly adequate, although it must be applied with great care. It is particularly important to include the effect of the external perturbation on the non-condensate, as well as  condensate phase and number fluctuations and various finite size effects, all of which are relevant in the JILA experiment. An advantage of the perturbative approach is that explicit expressions can be obtained for various response functions and this simplifies numerical calculations, allowing the effects of low-energy phonons and high-energy single-particles to be included in the same calculation.

The study of the dilute Bose gas has a long history dating back to the work of Bogoliubov in 1947 \cite{Bogoliubov47}, who described the excitations in terms of weakly-interacting quasiparticles. Beliaev extended the Bogoliubov approach by including interactions between quasiparticles using perturbation theory \cite{Beliaev58}. This approach was further developed by Hugenholtz and Pines \cite{Hugenholtz59} and by Popov and Fadeev \cite{Popov65a,Popov65b} who applied it to calculations at finite temperature. The Bogoliubov method was applied to inhomogeneous systems by de Gennes and Fetter \cite{deGennes66,Fetter72}, while the Beliaev-Popov formalism has recently been reexamined by Shi and Griffin \cite{Shi98} and extended to trapped gases by Fedichev and Shlyapnikov \cite{Fedichev98}. The excitation spectrum at zero and finite temperature was also studied by Mohling, Sirlin and Morita using many-body perturbation theory \cite{Mohling60}. These calculations have recently been extended to trapped gases by Morgan \cite{Morgan00}.

Dynamical properties of homogeneous and trapped condensates are well-described at low temperatures by the GPE \cite{GPE}. Condensate excitations can be studied by applying linear response theory to the GPE \cite{Ruprecht96,Edwards96}, reproducing the Bogoliubov quasiparticle results. Measurements of excitations at low-temperature are in good agreement with predictions based on this approach \cite{Jin96,Stringari96,Edwards96}. The GPE can be generalized to include the dynamic coupling to non-condensed atoms, which is important at finite temperature. Linear response theory has been applied to the generalized GPE by Giorgini \cite{Giorgini00}, and leads to results which are equivalent to the Beliaev-Popov formalism applied to trapped gases. Linear response theory at finite temperature has also been discussed by Minguzzi and Tosi \cite{Minguzzi97}, Guilleumas and Pitaevskii \cite{Guilleumas99} and by Rusch and Burnett \cite{Rusch99}, while the related dielectric formalism has been applied in this context by Bene and Sz\'{e}pfalusy \cite{Bene98} and Reidl \textit{et al.} \cite{Reidl00}. Recently, the GPE has been used to study the finite temperature dynamics of highly-occupied modes in Bose gases \cite{Davis01,Davis01b}, and a stochastic GPE has been developed to include the effect of quantum fluctuations \cite{Stoof01,Duine01,Gardiner02}.

Although low-temperature excitations can be studied straightforwardly using the GPE, finite-temperature measurements, particularly those made at JILA \cite{Jin97}, have proved much harder to explain. The starting point for many recent calculations of condensate excitations at finite temperature is the Hartree-Fock-Bogoliubov (HFB) theory, which has been summarized by Griffin \cite{Griffin96}. The full form of this theory is not gapless, however, and the HFB theory is therefore generally used in an approximation, often called the Popov approximation (HFBP), in which the anomalous (pair) average is neglected. The HFBP theory has been applied to trapped gases by a number of authors \cite{Hutchinson97,Dodd98,Reidl99}, and a gapless extension of the HFB formalism which includes the anomalous average has also been developed (GHFB) \cite{Proukakis98,Hutchinson98}. The various HFB approaches have been summarized recently in \cite{Hutchinson00}.

The application of the HFB approach to the 1997 JILA experiment \cite{Jin97} is particularly interesting. In this experiment the energies of the lowest-energy modes with axial angular momentum quantum numbers $m=2$ and $m=0$ were measured as a function of reduced temperature $t = T/T_c^0$, where $T$ is the absolute temperature and $T_c^0$ is the BEC critical temperature for an ideal gas. The $m=2$ mode was observed to shift downwards with $t$, while the $m=0$ mode underwent a sharp increase in energy at $t \sim 0.6$ towards the result expected in the non-interacting limit. HFBP calculations give good agreement with the JILA data for low temperatures but can not explain the results for $t>0.6$ \cite{Dodd98}. Good agreement for all $t$ for the $m=2$ mode was obtained using the GHFB theory, and also within the dielectric formalism \cite{Reidl00}. However, neither method was able to explain the upward shift of the $m=0$ mode, and an analytical calculation based on linearizing the generalized GPE also predicted downward shifts for both modes \cite{Giorgini00}.

A possible explanation for the behaviour of the $m=0$ mode has been given by Bijlsma, Al Khawaja and Stoof \cite{Bijlsma99,Khawaja00}. They emphasized the importance of the relative phase of condensate and non-condensate fluctuations and showed that the experimental results for $m = 0$ can be qualitatively explained by a shift from out-of-phase to in-phase oscillations at high temperature. Recently, Jackson and Zaremba \cite{Jackson02} obtained good agreement with the JILA experiment for both modes and also for the Oxford scissors mode experiment \cite{Jackson01}. Their calculations are based on the Zaremba-Nikuni-Griffin formalism \cite{Zaremba99} and use a generalized GPE for the condensate coupled to a non-condensate modelled by a semiclassical Boltzmann equation. Despite its successes, however, this approach neglects the phonon character of low-energy excitations as well as the anomalous average and all Beliaev processes. It is not a priori obvious that these are valid approximations. For example, in a weakly-interacting Bose condensate the anomalous average is as large as the contribution to the normal average from interactions \cite{Morgan00}, and it can be very significant near a Feshbach resonance \cite{Holland01,Kohler03}. The good agreement with the JILA results for the $m=2$ mode within the GHFB theory \cite{Hutchinson98} also suggests that the anomalous average can be important. Furthermore, Beliaev processes have been directly observed in a number of recent experiments \cite{Hodby01,Katz02,Bretin03,Mizushima03}. It is therefore important to develop a theory which includes these effects in a consistent manner, and we describe such a theory in this paper. An alternative kinetic theory which also includes these processes consistently has been developed by Walser \textit{et al.} \cite{Walser99,Walser00} and Wachter \textit{et al.} \cite{Wachter02}.

As well as measuring the energies of excitations as a function of temperature, experiments can also study the decay rates. The HFB theories do not describe the damping of excitations, however, because they treat the thermal cloud statically, whereas damping arises from the dynamical interaction between condensed and uncondensed atoms which removes energy from the condensate fluctuation. Damping of excitations has been studied theoretically by a number of authors (see for example \cite{Khawaja00,Guilleumas99,Bene98,Reidl00,Fedichev98,Jackson02,Giorgini98,Williams01,Fedichev98b,Pitaevskii97} among many others) and there is generally reasonable agreement with experimental measurements.

\subsection{Outline of the paper}

In this paper we present a theoretical description of condensate excitations which
has recently been used successfully to explain the JILA experimental results for
both the $m=2$ and $m=0$ modes \cite{Morgan03a}. Following Giorgini
\cite{Giorgini00}, we use linear response theory with a generalized GPE. We obtain
this equation using the number-conserving formalism of Castin and Dum
\cite{Castin98} with the result that it differs in detail from the more usual form
obtained via spontaneous symmetry breaking. The linear response treatment has the
advantage that it closely models the experimental procedure where excitations are
created by small modulations of the trap frequencies. In particular we can easily
include condensate phase and number fluctuations and the effect of the external
perturbation on the non-condensate, which were neglected in our earlier
time-independent approach \cite{Morgan00,Rusch00}.

This paper is organised as follows. In Sec.~\ref{sec_number_conserving_theory} we summarize the Castin-Dum number-conserving formalism and obtain equations of motion for the condensate and non-condensate. We also introduce the quasiparticle basis and discuss its relevant properties. In Sec.~\ref{UVRENORM} we briefly discuss the UV renormalization of the theory and outline its justification. In Sec.~\ref{sec_linear_response_theory} we develop the linear response formalism for a Bose gas, starting in Sec.~\ref{LINEAR_GPE} with the simplest approach based on the GPE and then considering in Sec.~\ref{LINEAR_GGPE} the higher order corrections which arise from the generalized GPE. Various issues related to the resulting theory are discussed in Sec.~\ref{sec_discussion}. In particular, in Sec.~\ref{sec_homogeneous} we consider the homogeneous limit which can be used to show that the theory is gapless and to obtain the conditions for its validity. In Sec.~\ref{sec_non-condensate_collisions} we show that non-condensate collisions are of higher order in the small parameter of the theory and can therefore be neglected for the purposes of this work. In Sec.~\ref{sec_dipole_mode} we show that the dipole modes are obtained correctly to within the order of the calculation. The appendices contain further details of the calculations.

We briefly comment on our notation, which is potentially complicated as a number of similar yet distinct quantities are required in the theory. We denote by a subscript `$0$' or `$2$' the relevant approximation to the condensate wavefunction, `$0$' for the GPE and `$2$' for the generalized GPE. Any time or frequency dependence is written explicitly (unless stated otherwise), so if none is given the corresponding quantity is calculated in equilibrium. Spatial dependence is usually written explicitly but is occasionally neglected when it is obvious. Most quantities in the theory depend on temperature, but to avoid overloading the notation we do not indicate this explicitly.

\section{Summary of time-dependent number-conserving formalism} \label{sec_number_conserving_theory}

Theoretical descriptions of dilute Bose condensates usually start from the assumption, first applied in this context by Bogoliubov \cite{Bogoliubov47}, that the $U(1)$ gauge symmetry of the Hamiltonian is spontaneously broken. While this procedure is undoubtedly useful there are some technical concerns with the violation of particle number-conservation that is involved. In fact, it has been shown by Girardeau and Arnowitt \cite{Girardeau59}, Gardiner \cite{Gardiner97}, and Castin and Dum \cite{Castin98} that symmetry breaking is not necessary and the Bogoliubov theory and higher order calculations can be obtained via a number-conserving approach. In this paper we use the time-dependent formalism of Castin and Dum \cite{Castin98}, with minor extensions to finite temperature, to obtain the equations of motion for the condensate and non-condensate. At leading order there is no difference between this approach and a symmetry-breaking formalism. At higher order, however, the number-conserving approach describes various finite size effects which are not contained in symmetry-breaking calculations. We found in \cite{Morgan03a} that these effects are important for explaining the JILA experimental results within this formalism, although it is an open question to what extent they are significant more generally.

We have chosen to use an explicitly number-conserving approach rather than a symmetry-breaking treatment because we will make regular use of the Bogoliubov quasiparticle basis to expand various wavefunctions. It is therefore important to have a well-defined, complete basis set and in particular to deal correctly with the zero-energy condensate modes which appear. This is simplest in the number-conserving approach where these states occur naturally and are treated on the same footing as those with non-zero energy. In particular, there is no `missing eigenvector' in this approach and issues related to the orthogonality of excitations to the condensate are dealt with explicitly.

In this section we summarize the Castin-Dum number-conserving formalism. The original derivation given in \cite{Castin98} was restricted to near zero temperatures because it was assumed that the condensate population $N_0$ was approximately equal to the fixed total number of atoms $N$. The changes required to adapt the approach to perturbative calculations at finite temperature are described in more detail elsewhere \cite{Morgan03d}. In this section we quote the equations that result and discuss the properties that are relevant to the current work.

\subsection{Equations of motion}

We start from the usual Heisenberg equation of motion for the field operator $\hat{\Psi}\rt$ which annihilates a particle at point $\bfr$ and time $t$
\begin{align}
i \hbar \frac{\partial}{\partial t}\hat{\Psi}\rt &= \hsp\rt
\hat{\Psi}\rt \nonumber \\
&\phantom{=} + U_0 \hat{\Psi}^{\dagger}\rt \hat{\Psi}\rt \hat{\Psi}\rt,
\label{Op_GPE}
\end{align}
where $\hsp\rt = -(\hbar^2/2m) \nabla^2 + V \rt$ is the single-particle Hamiltonian containing the kinetic energy and any external potential. For simplicity we have assumed binary interactions between the particles characterized by an energy-independent contact potential, $V_{\text{bin}}(\bfr-\mathbf{r'}) = U_0 \delta(\bfr-\mathbf{r'})$, where $U_0 = 4\pi\hbar^2a_s/m$ and $a_s$ is the s-wave scattering length. This is the standard approximation for three-dimensional, dilute Bose gases. As is well known, however, it leads to ultra-violet divergences which must be removed by renormalizing various quantities which appear in the subsequent development of the theory. This procedure is well understood and has been discussed by a number of authors (see for example \cite{Huang_StatMech,Stoof93,Morgan00,Olshanii02} among many others). It can be rigorously justified and we give a brief outline of the relevant arguments in Sec.~\ref{UVRENORM}.

Following the definition of Penrose and Onsager \cite{Penrose56}, a condensate is said to be present if the one-body density operator $\rho_1 (\bfr, \mathbf{r'},t) = \langle \hat{\Psi}^{\dagger}(\mathbf{r'},t) \hat{\Psi}\rt \rangle$ has an eigenvector with a macroscopic eigenvalue much larger than all others. Here $\langle \ldots \rangle$ denotes a statistical average in the appropriate ensemble. The exact condensate wavefunction $\Phi_{\text{ex}} \rt$ (normalized to one) and population $N_0(t)$ therefore satisfy the equation
\beq
\intdrp \rho_1 (\bfr, \mathbf{r'},t)\Phi_{\text{ex}} \rpt = N_0(t)\Phi_{\text{ex}} \rt,
\label{cond_def}
\eeq
with $N_0(t) \gg 1$. We can write the full field operator in terms of condensate and non-condensate operators as
\beq
\hat{\Psi}\rt = \Phi_{\text{ex}} \rt \hat{a}_0 (t) + \delta\hat{\Psi}\rt,
\eeq
where the condensate annihilation operator $\hat{a}_0 (t)$ is defined by projecting $\hat{\Psi}\rt$ onto $\Phi_{\text{ex}} \rt$, while $\delta\hat{\Psi}\rt$ is obtained via the orthogonal projection
\begin{align}
\hat{a}_0 (t) &= \intdr \Phi_{\text{ex}}^* \rt \hat{\Psi}\rt, \label{def_a0}\\
\delta\hat{\Psi}\rt &= \intdrp Q_{\text{ex}}(\bfr,\mathbf{r'},t) \hat{\Psi} (\mathbf{r'},t) \equiv \hat{Q}_{\text{ex}}(t)\hat{\Psi}\rt. \label{Qhat}
\end{align}
The orthogonal projector $Q_{\text{ex}}(\bfr,\mathbf{r'},t)$ is defined by
\beq
Q_{\text{ex}}(\bfr,\mathbf{r'},t) = \delta (\bfr-\mathbf{r'}) - \Phi_{\text{ex}} \rt \Phi_{\text{ex}}^*(\mathbf{r'},t),
\label{def_Q}
\eeq
and the final part of Eq.~(\ref{Qhat}) defines a convenient shorthand notation for the action of this operator. The mean condensate population is given as usual by $N_0(t) = \langle \hat{a}_0^{\dagger} (t) \hat{a}_0 (t)\rangle$.

Eq.~(\ref{cond_def}) leads directly to
\beq
\langle \hat{a}_0^{\dagger} (t) \delta\hat{\Psi}\rt \rangle = 0,
\label{av_a0delta}
\eeq
which is the defining relation from which equations for $\Phi_{\text{ex}} \rt$ and $\delta\hat{\Psi}\rt$ follow \cite{Castin98}. The equation of motion for $\hat{a}_0^{\dagger} (t) \delta\hat{\Psi}\rt$ can be written as
\beq
i \hbar \frac{\partial}{\partial t} \left [ \hat{a}_0^{\dagger} (t) \delta\hat{\Psi}\rt \right ] = \intdrp \sum_{k=0}^{4} \hat{R}_k \rrt,
\label{a0delta_eqn}
\eeq
where the quantities $\hat{R}_k \rrt$ are operators containing $k$ factors of the non-condensate field $\delta\hat{\Psi}$. In the presence of a condensate we can therefore assume that the effect of $\hat{R}_{k+1}$ is small compared to $\hat{R}_k$ and use this as a means of developing a systematic approximation scheme. Including progressively more of the $\hat{R}_k$ leads to a sequence of improving approximations to the exact condensate wavefunction, which we denote by $\Phi_0 \rt$, $\Phi_1 \rt$, $\Phi_2 \rt$ etc.

The leading order theory is obtained by neglecting all terms except $\hat{R}_0$, which gives an approximate equation of motion for the condensate wavefunction $\Phi_{\text{ex}} \rt \rightarrow \Phi_0 \rt$ which is just the usual time-dependent GPE
\begin{multline}
i \hbar \frac{\partial}{\partial t} \Phi_0 \rt = \left [ \hsp\rt- \lambda_0(t) \right ]\Phi_0 \rt \\
+ N_0U_0 | \Phi_0 \rt|^2 \Phi_0 \rt.
\label{TDGPE}
\end{multline}
Here $\Phi_0 \rt$ is normalized to one and $\lambda_0(t)$ is an arbitrary, real function of time which can be chosen at will to adjust the global phase of the condensate wavefunction. At this level of approximation the mean condensate population $N_0$ is independent of time.

Improved approximations for the condensate wavefunction can be obtained by including more of the operators $\hat{R}_k$. It turns out that $\langle \hat{R}_1 \rangle = 0$, so $\Phi_1 \rt = \Phi_0 \rt$. The next approximation therefore comes from including the effect of $\hat{R}_2$, $\Phi_{\text{ex}} \rt \rightarrow \Phi_2 \rt$. The resulting equation of motion is the generalized GPE (GGPE)
\begin{align}
 i \hbar \frac{\partial}{\partial t} \Phi_2 \rt &= \left[ \hsp\rt - \lambda_2(t)\right ] \Phi_2 \rt \nonumber \\
&\phantom{=}+ \bigl ( N_0(t) + \Delta N_0 \bigr ) U_0|\Phi_2 \rt|^2 \Phi_2\rt \nonumber \\
&\phantom{=}+ 2 U_0 \nt\rt\Phi_2\rt + U_0 \mt \rt \Phi_2^* \rt \nonumber\\
&\phantom{=}- f\rt, \label{TDGGPE}
\end{align}
where the function $f\rt$ is defined by
\begin{multline}
f \rt = \intdrp U_0 |\Phi_2 \rpt|^2 \Bigl [\Bigr. \Phi_2 \rpt \nt \rrt\\
 + \Phi_2^* \rpt \mt \rrt \Bigl. \Bigr ],
\label{def_f}
\end{multline}
and $\Delta N_0$ is defined by
\beq
\Delta N_0 = \frac{\langle \hat{N}_0^2 \rangle-\langle \hat{N}_0 \rangle^2}{N_0} -1.
\label{DN_0}
\eeq

The contribution to the GGPE from $f\rt$ was first obtained by Castin and Dum \cite{Castin98} and does not appear in symmetry-breaking approaches to the theory of BEC. We will see later that it is related to the strict enforcement of orthogonality between the condensate and non-condensate. The contribution from statistical fluctuations in the condensate population ($\Delta N_0$) is new and comes from the extension of the Castin-Dum analysis to finite temperature \cite{Morgan03d}. In practice, these fluctuations are negligible, both because they are intrinsically small even for condensates containing only a few thousand atoms, and also because the properties of the condensate are insensitive to the exact number of atoms when $N_0$ is large. For completeness, however, we retain this term in the GGPE during the development of the theory. An expression for $\Delta N_0$ in terms of Bogoliubov quasiparticles is given in Appendix~\ref{app_condensate_statistical_fluctuations}.

The off-diagonal non-condensate density matrix $\nt\rrt$ and anomalous average $\mt\rrt$ which appear in Eqs.~(\ref{TDGGPE}) and (\ref{def_f}) are defined by
\begin{align}
\nt \rrt &= \left \langle \hat{\Lambda}^{\dagger}\rpt \hat{\Lambda}\rt \right \rangle,\\
\mt \rrt &= \left \langle \hat{\Lambda}\rpt \hat{\Lambda}\rt \right \rangle,
\end{align}
where $\hat{\Lambda}\rt$ is the number-conserving, non-condensate field operator \cite{Gardiner97,Castin98,Girardeau59}
\beq
\hat{\Lambda} \rt \equiv \frac{1}{\sqrt{\hat{N}_0}} \hat{a}_0^{\dagger} (t) \delta\hat{\Psi}\rt.
\eeq
For the diagonal parts of the non-condensate density and anomalous average we use the notation $\nt\rt \equiv \nt (\bfr,\bfr,t)$ and $\mt\rt \equiv \mt (\bfr,\bfr,t)$.

At this level of approximation, the mean condensate population $N_0(t)$ is no longer independent of time. The condensate population is determined by the non-condensate population $N_{\text{nc}}(t) = \intdr \nt\rt$ because the total atom number is fixed
\beq
N_0(t)+N_{\text{nc}}(t) = N \Rightarrow \frac{d N_0}{dt} = -\frac{d N_{\text{nc}}}{dt}.
\label{dN0_dt}
\eeq
In fact we use this relation to \emph{determine} the condensate population in terms of $N$ and the temperature $T$. The parameter $N_0(t)$ is therefore just a shorthand notation for $N-N_{\text{nc}}(t)$. Equation~(\ref{dN0_dt}) shows that the evolution of $N_0(t)$ depends quadratically on the non-condensate field operator. This means that its time-dependence is only relevant when we consider the effect of $\hat{R}_2$ on the condensate wavefunction, i.e. in the GGPE of Eq.~(\ref{TDGGPE}), and at leading order we can assume that $N_0$ is independent of time.

As a consequence of the time-dependence of $N_0(t)$, the parameter $\lambda_2(t)$ in Eq.~(\ref{TDGGPE}) has an imaginary part $\lambda_2^{i}(t)$ in order to keep the normalization of the wavefunction constant, $\intdr |\Phi_2 \rt|^2 = 1$. This condition gives
\begin{align}
\lambda_2^{i}(t) &= \frac{1}{2i}\intdr \left [ U_0 \mt\rt \Phi_2^{*2}\rt - c.c. \right ] \nonumber \\
&= \frac{\hbar}{2N_0}\frac{dN_0}{dt},
\label{lambdaI}
\end{align}
where c.c. stands for `complex conjugate' and the last line follows from the equation of motion for $N_0(t)$ to the order of this calculation. The imaginary part of $\lambda_2(t)$ is therefore determined by the equations of motion and only the real part is arbitrary (as before, this corresponds to choosing the global phase of the condensate wavefunction). Eq.~(\ref{lambdaI}) shows that at this level of approximation, the dominant processes changing the condensate population involve pair excitation into and out of the condensate.

The GGPE must be combined with an equation of motion for the non-condensate field operator $\hat{\Lambda}\rt$. Correct to the order of this calculation, this is given by
\beq
i \hbar \frac{\partial}{\partial t} \bpm \hat{\Lambda}\rt \\ \hat{\Lambda}^{\dagger}\rt \epm = \mathcal{L}_2\rt \bpm \hat{\Lambda}\rt \\ \hat{\Lambda}^{\dagger}\rt \epm,
\label{TDLambda}
\eeq
where the matrix operator $\mathcal{L}_2\rt$ is defined by
\beq
\mathcal{L}_2\rt = \bpm \hat{L}_2\rt & \hat{M}_2\rt \\ -\hat{M}_2^*\rt & -\hat{L}_2^*\rt \epm,
\label{bigL2rt}
\eeq
\begin{align}
\hat{L}_2\rt = \hsp\rt - \lambda_2(t) + N_0U_0 |\Phi_2\rt|^2& \nonumber \\
+ N_0U_0 \hat{Q}_2(t)|\Phi_2\rt|^2\hat{Q}_2(t)&,\label{L2rt}
\end{align}
\beq
\hat{M}_2\rt = N_0 U_0\hat{Q}_2(t)\Phi_2^2\rt\hat{Q}_2^*(t).\label{M2rt}
\eeq
The notation $\hat{Q}_2(t)|\Phi_2\rt|^2\hat{Q}_2(t)$ means, for example,
\begin{multline}
\Bigl [ \hat{Q}_2(t)|\Phi_2\rt|^2\hat{Q}_2(t) \Bigr ] \hat{\Lambda}\rt \equiv \\
\iint \!\! d^3\bfrp d^3\mathbf{r''} \, Q_2\rrt |\Phi_2\rpt|^2 \\ \times Q_2(\mathbf{r'},\mathbf{r''},t)\hat{\Lambda}(\mathbf{r''},t),
\end{multline}
where $Q_2\rrt$ is defined as in Eq.~(\ref{def_Q}) with $\Phi_{\text{ex}}\rt \rightarrow \Phi_{2}\rt$. Equation~(\ref{TDLambda}) differs from the form one would obtain in a symmetry-breaking approach by the appearance of the orthogonal projectors $\hat{Q}_2$. These preserve orthogonality between the time-dependent condensate and non-condensate and are unique to the number-conserving formalism. A subtle issue concerns the appearance of $\Phi_2\rt$ rather than $\Phi_0\rt$ in Eqs.~(\ref{bigL2rt})-(\ref{M2rt}). This is necessary to obtain sensible expressions for energy shifts and is discussed further in Appendix~\ref{app_phi2_in_L}.

The GPE of Eq.~(\ref{TDGPE}), its generalization of Eq.~(\ref{TDGGPE}) and Eq.~(\ref{TDLambda}) for the non-condensate (together with the definitions of all relevant quantities) define the time-dependent, number-conserving, mean-field theory we use in this paper. Before applying these equations, however, we first describe the quasiparticle basis obtained from the eigenstate solutions of Eq.~(\ref{TDLambda}) and discuss some of its properties.

\subsection{Bogoliubov Quasiparticles}

Castin and Dum \cite{Castin98} have shown that the solutions of Eq.~(\ref{TDLambda}) can be written in terms of a basis expansion with time-independent quasiparticle annihilation operators and time-dependent eigenmodes. This is very useful in the subsequent development of the theory. The static case defines a convenient basis for the calculation while the dynamic case gives expressions for the full off-diagonal time-dependent non-condensate density and anomalous average.

\subsubsection{Static case}

We consider first the situation that the condensate wavefunction satisfies the static limit of the ordinary GPE of Eq.~(\ref{TDGPE}). If the external potential is independent of time, we can find solutions $\Phi_0 \rt = \cond$ which satisfy
\beq
\left [ \hsp(\bfr) - \lambda_0 + N_0U_0 |\cond|^2 \right ]  \cond = 0,
\label{TIGPE}
\eeq
where $\lambda_0$ (the condensate eigenvalue) is a constant, approximately equal to the chemical potential of the system, and the single-particle Hamiltonian $\hsp(\bfr)$ contains the kinetic energy and any static trap potential
\beq
\hsp(\bfr) = -\frac{\hbar^2 \nabla^2}{2m} + V_{\text{trap}}(\bfr).
\label{hsp}
\eeq

We now consider the matrix operator $\mathcal{L}_2\rt$ for this static case with the replacements $\Phi_2 \rt \rightarrow \cond$, $\lambda_2(t)=\lambda_0$. This gives the operator $\mathcal{L}_0(\bfr)$, defined by making these changes in Eqs.~(\ref{bigL2rt})-(\ref{M2rt})
\beq
\mathcal{L}_0(\bfr) = \bpm \hat{L}_0(\bfr) & \hat{M}_0(\bfr) \\ -\hat{M}^*_0(\bfr) & -\hat{L}^*_0(\bfr) \epm,
\label{bigL0r}
\eeq
\begin{align}
\hat{L}_0(\bfr) = \hsp(\bfr) - \lambda_0 +N_0U_0|\Phi_0(\bfr)|^2 & \nonumber\\
+ \hat{Q}_0|\Phi_0(\bfr)|^2\hat{Q}_0,&
\label{L0r}
\end{align}
\beq
\hat{M}_0(\bfr) = N_0 U_0\hat{Q} _0\Phi_0^2(\bfr)\hat{Q} _0^*.\label{M0r}
\eeq
The eigenstates and eigenvalues of $\mathcal{L}_0(\bfr)$ are the solutions of the Bogoliubov-de Gennes (BdG) equations
\beq
\mathcal{L}_0(\bfr) \mathcal{X}_i(\bfr) = \epsilon_i \mathcal{X}_i(\bfr),
\label{LBdG_matrix}
\eeq
where the two-component spinor $\mathcal{X}_i(\bfr)$ is defined in terms of quasiparticle wavefunctions $u_i(\bfr)$ and $v_i(\bfr)$ by
\beq
\mathcal{X}_i(\bfr) = \bpm u_i(\bfr) \\ v_i(\bfr) \epm,
\eeq
and the $\epsilon_i$ are the quasiparticle energies. These solutions define the static quasiparticle basis we use in this paper. At this level of approximation, the theory is simply the Bogoliubov theory with orthogonalized quasiparticle wavefunctions.

The eigenvectors $\mathcal{X}_i(\bfr)$ have a number of important properties that we will need later \cite{Castin98,Blaizot_Ripka}. They obey a generalized orthonormality relation which can be summarized by
\beq
\intdr \mathcal{X}_i^{\dagger}(\bfr) \sigma_3 \mathcal{X}_j(\bfr) = \chi_i \delta_{ij},
\label{orthosym}
\eeq
where $\chi_i = \pm 1$ is the norm of $\mathcal{X}_i(\bfr)$ and $\sigma_3 = \text{\footnotesize $\bpm 1 & 0 \\ 0 & -1 \epm$}$ is the third Pauli matrix. If $\mathcal{X}_i(\bfr)$ is an eigenvector with eigenvalue $\epsilon_i$ and norm $\chi_i$ then
\beq
\mathcal{X}_{-i}(\bfr) = \sigma_1 \mathcal{X}_{i}^*(\bfr) = \bpm v_i^*(\bfr) \\ u_i^*(\bfr) \epm,
\label{X_minus_i}
\eeq
is an eigenvector with eigenvalue $-\epsilon_i$ and norm $-\chi_i$ ($\sigma_1 = \text{\footnotesize $\bpm 0 & 1 \\ 1 & 0 \epm$}$ is the first Pauli matrix). If $\cond$ is a stable ground state solution of the GPE, then states with positive norm have positive or zero energy while states with negative norm have negative or zero energy. This is not true, however, if $\cond$ contains a vortex. In this case there exist anomalous modes with positive norm and negative energy which have been the source of much recent discussion (see for example \cite{Dodd97,Isoshima99,Svidzinksy00,Feder01,Virtanen01}).

The operator $\mathcal{L}_0(\bfr)$ is diagonalizable so its eigenvectors form a complete set and there are no missing modes in this formalism. In particular, there are two zero-energy solutions to Eq.~(\ref{LBdG_matrix}) which we denote by $\pm0$, according as their norm is $\pm1$. These two states involve the condensate wavefunction and are given by
\begin{align}
\mathcal{X}_{+0}(\bfr) &= \bpm u_{+0}(\bfr) \\ v_{+0}(\bfr) \epm = \bpm \cond \\ 0 \epm, \label{X0plus}\\
\mathcal{X}_{-0}(\bfr) &= \bpm u_{-0}(\bfr) \\ v_{-0}(\bfr) \epm = \bpm 0 \\ \ccond \epm.
\end{align}
An advantage of the number-conserving formalism, is that these two condensate modes obey the same relations as states with non-zero energy and can therefore be treated on exactly the same footing. In particular, they are related by the same interchange of $u$ and $v$ functions that applies to other pairs of positive/negative energy states and they obey the orthonormality condition, which in this case simply expresses the orthogonality of modes with non-zero energy to the condensate
\beq
\intdr \ccond u_i(\bfr) = \intdr \cond v_i(\bfr) = 0, \quad (i \neq \pm 0).
\eeq

It turns out to be convenient to form a linear combination of the two condensate modes to obtain vectors describing changes in the condensate norm and phase
\beq
\mathcal{X}_{N_0}(\bfr) = \bpm \cond \\ \ccond \epm, \quad \mathcal{X}_{\theta_0}(\bfr) = \bpm \cond \\ -\ccond \epm.
\label{def_Xnorm_Xphase}
\eeq
These states satisfy the generalized orthogonality of Eq.~(\ref{orthosym}) with all other states but have zero norm. They can be projected out of any expansion using the result
\beq
\frac{1}{2}\intdr \mathcal{X}_{N_0}^{\dagger}(\bfr) \sigma_3 \mathcal{X}_{\theta_0}(\bfr) = 1.
\label{overlap_Xnorm_Xphase}
\eeq 

The completeness of the basis vectors $\mathcal{X}_i(\bfr)$ gives the decomposition of unity
\beq
\sum_{i} \chi_i\mathcal{X}_i(\bfr)\sigma_3\mathcal{X}_i^{T}(\bfrp)= \delta(\bfr - \bfrp) \bpm 1 & 0 \\ 0 & 1 \epm,
\label{completeness}
\eeq
where the sum is over all states. The fact that these vectors form a complete set of states is very important for our subsequent use of them as a basis to expand condensate and non-condensate fluctuations.

\subsubsection{Dynamic case}

In the dynamic case, the non-condensate field operator $\hat{\Lambda}\rt$ can be written in terms of time-independent quasiparticle creation and annihilation operators (respectively $\hat{\beta}_i^{\dagger}$, $\hat{\beta}_i$) and time-dependent eigenmodes \cite{Castin98}. Since $\hat{\Lambda}\rt$ is orthogonal to the condensate we can write
\beq
\bpm \hat{\Lambda}\rt \\ \hat{\Lambda}^{\dagger}\rt \epm = \sum_{i>0} \mathcal{X}_i\rt\hat{\beta}_i + \mathcal{X}_{-i}\rt\hat{\beta}_i^{\dagger},
\label{Lqpexp}
\eeq
where the summation is over all non-condensate modes with positive norm, $\chi_i = +1$. The spinors $\mathcal{X}_i\rt$ are defined in terms of time-dependent Bogoliubov wavefunctions $u_{i}\rt$ and $v_{i}\rt$ by
\beq
\mathcal{X}_i\rt = \bpm u_{i}\rt \\ v_{i}\rt \epm,\quad \mathcal{X}_{-i}\rt = \bpm v_{i}^{*}\rt \\ u_{i}^{*}\rt \epm,
\label{def_uvrt}
\eeq
and evolve according to
\beq
i \hbar \frac{\partial}{\partial t}\mathcal{X}_i\rt  = \mathcal{L}_2\rt \mathcal{X}_i\rt.
\label{uvt}
\eeq
In the equilibrium case, the spinors $\mathcal{X}_i\rt$ are related to the static basis functions defined above by $\mathcal{X}_i\rt = \mathcal{X}_i(\bfr)e^{-i\epsilon_it/\hbar}$.

In this formulation, the quasiparticle creation and annihilation operators are independent of time
\beq
\frac{d\hat{\beta}_i^{\dagger}}{d t} = \frac{d\hat{\beta}_i}{d t} = 0.
\label{dBeta_dt}
\eeq
It follows from the orthonormality relations and the definition of $\hat{\Lambda} \rt$ that the quasiparticle operators obey bosonic commutation relations (neglecting the subspace with $N_0 = 0$). We note that Eq.~(\ref{uvt}) [with $\Phi_2\rt \rightarrow \Phi_0\rt$ in $\mathcal{L}_2\rt$] also gives the evolution of the condensate to leading order because in this case it reduces to the GPE of Eq.~(\ref{TDGPE}). This ensures preservation of orthogonality between the condensate and excited states to the order of the calculation in the time-dependent case.

Equations~(\ref{Lqpexp}) and (\ref{uvt}) provide the full equation of motion for the non-condensate from which $\nt\rrt$ and $\mt\rrt$ can be constructed. A representation of these quantities in terms of the spinors $\mathcal{X}_i\rt$ is given in Appendix~\ref{app_dynamic_shifts}, Eqs.~(\ref{app_def_R})-(\ref{app_Ri}) and is very convenient for calculations. They can also be written in a compact form using the time-dependent quasiparticle wavefunctions $u_i\rt$ and $v_i\rt$ as
\begin{align}
\nt \rrt = \sum_{i>0}  &u_i\rt u_i^*\rpt N_i \nonumber \\ + &v_i^*\rt v_i\rpt (N_i + 1),\label{nt_qp}
\end{align}
\begin{align}
\mt \rrt = \sum_{i>0} &u_i\rt v_i^*\rpt N_i \nonumber\\ + &v_i^*\rt u_i\rpt (N_i + 1),\label{mt_qp}
\end{align}
where we have assumed that $\langle \hat{\beta}_i^{\dagger} \hat{\beta}_j \rangle = N_i \delta_{ij}$ and $\langle \hat{\beta}_i \hat{\beta}_j \rangle = 0$, which are appropriate for non-interacting quasiparticles. As a consequence of Eq.~(\ref{dBeta_dt}), the quasiparticle populations $\{ N_i \}$ are independent of time, as has also been shown by Walser \textit{et al.} \cite{Walser00}. Substituting Eqs.~(\ref{nt_qp}) and (\ref{mt_qp}) into Eq.~(\ref{def_f}) gives the form of $f\rt$ quoted in \cite{Morgan03a}
\beq
f\rt = \frac{1}{N_0}\sum_{i} c_i^*(t)N_i u_i\rt + c_i(t)(N_i+1)v_i^*\rt,
\label{frt}
\eeq
where the scalars $c_i(t)$ are defined by
\begin{align}
c_i(t) = \intdr N_0U_0|\Phi_2\rt|^2 \Bigl [ \Bigr. &\Phi_2^*\rt u_i\rt \nonumber \\
&+ \Phi_2\rt v_i\rt \Bigl. \Bigr ].
\label{c_coeffs_t}
\end{align}

We will be interested in the case that the quasiparticle populations $\{ N_i \}$ correspond to thermal equilibrium so that they are given by the Bose-Einstein distribution
\beq
N_i(\epsilon_i,\beta) = \frac{1}{e^{\beta (\epsilon_i-[\mu-\lambda_0])}-1},
\label{N_i_kT}
\eeq
where $\mu$ is the chemical potential, $\beta = 1/k_{\text{B}}T$ and $k_{\text{B}}$ is Boltzmann's constant. The difference between $\mu$ and $\lambda_0$ is of order $1/N_0$, but for systems with a finite number of particles this difference must be accounted for otherwise it is impossible to find equilibrium solutions that satisfy the constraint on $N$ near the critical temperature (and beyond). We note that most quantities in the theory depend on temperature via their dependence on these populations [either explicitly as in $\nt\rrt$ and $\mt\rrt$ or implicitly as in $\Phi_2\rt$ and $N_0(t)$].

\section{Ultra-violet renormalization}\label{UVRENORM}

The GGPE of Eq.~(\ref{TDGGPE}) contains the diagonal part of the anomalous average $\mt\rt$ which is ultra-violet (UV) divergent. The reason and cure for this problem are well-known and we give a brief summary of the argument in this section.

The UV divergence in the anomalous average arises because of the use of the contact potential approximation to describe particle interactions. A truly ab initio theory would have to start by describing these interactions using the actual interatomic potential. In contrast, the contact potential is the zero-energy, zero-momentum limit of the homogeneous, two-body T-matrix describing the scattering of two particles in a vacuum. Detailed arguments which show formally how the bare interaction potential is replaced by the two-body T-matrix can be found in \cite{Morgan00,Rusch99,Proukakis98b,Kohler02,Wachter02,Stoof93,Bijlsma97}. This replacement naturally involves a renormalization of higher order terms.

Rather than rederive these results, we simply introduce the T-matrix into the theory from the start, which is why the contact potential appears in Eq.~(\ref{Op_GPE}) for the full quantum field operator. This does mean, however, that we have implicitly included at the outset various physical effects (multiple two-body scattering) which must also appear in the many-body treatment. To avoid double counting these effects we must identify where they appear in higher order calculations and explicitly subtract them in the appropriate approximation. Since the leading order contribution from interactions is the nonlinear part of the GPE, this is the first term requiring renormalization. In the GGPE, the interaction strength $U_0$ in this term must therefore be replaced by the expression \cite{Giorgini00,Morgan00}
\beq
U_0 \rightarrow U_0 + \Delta U_0,\\
\eeq
where $\Delta U_0$ is the second-order approximation to the interaction strength as calculated from the Lippmann-Schwinger equation
\beq
\Delta U_0 = U_0^2 \int \frac{d^3{\bf k}}{(2\pi)^3} \frac{1}{2(\hbar^2k^2/2m)}.
\label{DeltaU0}
\eeq

We note that this correction is appropriate even for gases in a trapping potential despite the fact that it involves an integral in momentum space. This is because $U_0$ corresponds to the homogeneous T-matrix, so for a trapped gas two-body collisions have been implicitly included via a local density approximation and must therefore be subtracted in the same manner. The integral in Eq.~(\ref{DeltaU0}) is itself UV divergent, but this cancels with the divergence in the anomalous average so that the combination is finite [see Eq.~(\ref{rmt})].

We therefore have a correction term in the GGPE which is $N_0 \Delta U_0 |\Phi_2 \rt|^2\Phi_2 \rt$. This correction can be grouped with the diagonal part of the anomalous average, leading to a finite, renormalized version of this quantity defined by
\beq
\mt^{\text{R}}\rt = \mt\rt + N_0 \frac{\Delta U_0}{U_0} \Phi_2^2\rt.
\label{rmt}
\eeq
The renormalized, generalized GPE is therefore given by Eq.~(\ref{TDGGPE}) with $\mt\rt $ replaced by $\mt^{\text{R}}\rt$.

We note that this procedure only affects the diagonal part of the anomalous average. The off-diagonal terms $\mt\rrt$ with $\bfr \neq \bfrp$ do not require any UV renormalization and are unaffected. In particular, the contribution to $f\rt$ in Eq.~(\ref{def_f}) from $\mt\rrt$ does not contain any UV renormalization.

\section{Theory of linear response at zero and finite temperature} \label{sec_linear_response_theory}

In this section we apply the number-conserving, mean-field theory described above to consider the linear response of a Bose-condensed system at zero and finite temperature. We closely follow the method of Giorgini \cite{Giorgini00}, although the calculation differs in detail because it contains effects due to the orthogonal projector $\hat{Q}$ and the term $f\rt$ in Eq.~(\ref{def_f}) which are absent in the more usual symmetry-breaking formalism. We have found numerically that these terms can have a significant effect for the parameters of the JILA experiment \cite{Jin97,Morgan03a,Morgan03d}. We also explicitly include the effect of
the external perturbation on the non-condensate. This is generally neglected in analytical calculations but is vital for an understanding of the JILA experiment \cite{Jin97,Morgan03a} as well as for an accurate description of the dipole modes.

We start in Sec.~\ref{LINEAR_GPE} with the simplest theory obtained by using only the GPE, which is generally sufficient at zero temperature. In Sec.~\ref{LINEAR_GGPE} we then consider the corrections which arise from the GGPE. These are important at finite temperature but they also lead to changes in the zero temperature results. Our results apply to the collisionless limit in which the characteristic time scale for quasiparticle collisions is long compared with the chemical potential. More formally, the theory is based on a systematic expansion in a small parameter which for a homogeneous system of volume $V$ is given by \cite{Fedichev98}
\beq
\begin{alignedat}{2}
(na_s^3)^{1/2} \ll &1, & \qquad &(T = 0), \\
\Bigl ( \frac{k_{\text{B}}T}{n_0U_0} \Bigr ) (n_0a_s^3)^{1/2} \ll &1, & \qquad &\left (\frac{k_{\text{B}}T}{n_0U_0} \gg 1 \right ),
\end{alignedat}
\label{validity}
\eeq
where $n_0=N_0/V$ and $n = N/V$ are the condensate and total densities respectively. The origin of these expressions is discussed further in Secs.~\ref{sec_homogeneous} and \ref{sec_non-condensate_collisions}.

We note that the theory is not an expansion to second order in the coupling constant $U_0$ as stated in \cite{Giorgini00}. This can be seen from the fact that terms of order $U_0^2$, corresponding to collisions within the non-condensate, are neglected. Instead, we calculate the leading order corrections to the Bogoliubov theory in the small parameter given above. We show in Sec.~\ref{sec_non-condensate_collisions} that non-condensate collisions are of higher order in this parameter and can therefore be neglected for the purposes of this calculation. A kinetic theory which includes all terms of order $U_0^2$ has been given by Walser \textit{et al.} \cite{Walser99} and, in the perturbative limit, by Rusch and Burnett \cite{Rusch99}.

We note that at finite temperature the small parameter of the theory is proportional to $1/\sqrt{n_0}$ and hence becomes large close to the BEC phase transition. Setting this parameter equal to one gives an estimate for the boundary of the critical region which is closely related to the Ginzburg criterion for the failure of mean-field theory \cite{Huang_StatMech}. Inside the critical region perturbation theory will fail, but for a dilute gas this region is very narrow and our approach is valid for a wide temperature range. If Eq.~(\ref{validity}) is used to estimate the validity of the theory for a trapped gas by replacing $n_0$ with $n_0(\bfr)$ via a local density approximation, the small parameter becomes spatially dependent and diverges at the edge of the condensate. This is primarily a consequence of the crudity of the estimate but perhaps indicates that the theory may be more successful for modes localized near the centre of the condensate than around its edge. Evaluating the parameter using the central condensate density probably provides a more useful diagnostic, but the validity of the theory in this case is best checked a posteriori simply by seeing if the predicted energy shifts and widths are small compared to the unperturbed values.

\subsection{GPE theory} \label{LINEAR_GPE}

The excitations of a Bose gas can by studied by considering the linear response of the system to an infinitesimal time-dependent external perturbation $P\rt$ \cite{Ruprecht96,Edwards96,Giorgini00}. The leading order theory (Bogoliubov theory) is obtained using just the GPE of Eq.~(\ref{TDGPE}). We consider the case that the external potential $V\rt$ can be written as a sum of two terms, a static trap potential $V_{\text{trap}}(\bfr)$ (if any) and the perturbation $P\rt$, both of which are real functions. We therefore write the GPE of Eq.~(\ref{TDGPE}) as
\begin{align}
 i \hbar \frac{\partial}{\partial t} \Phi_0 \rt = \left [ \hsp(\bfr) + P\rt - \lambda_0(t) \right ]  \Phi_0 \rt& \nonumber\\
 + N_0U_0 | \Phi_0 \rt|^2 \Phi_0 \rt&,
 \label{TDGPE2}
\end{align}
where $\hsp(\bfr)$ is given in Eq.~(\ref{hsp}) and $N_0$ is independent of time.

In the static case where $P\rt=0$, the GPE has time-independent solutions $\Phi_0 \rt = \cond$ given by Eq.~(\ref{TIGPE}). We will consider the situation where a condensate has been formed at low temperature and has settled into a solution of this equation. We note that the formalism we will develop applies in principle to any such solution and not just the ground state. In particular the results can be applied to vortex states, and for this reason we have been careful not to assume that $\cond$ is real.

When the external perturbation $P\rt$ is applied, the system responds with a time-dependent oscillation of the mean field
\beq
\Phi_0 \rt = \cond + \delta \Phi_0 \rt.
\eeq
In general $\delta \Phi_0 \rt$ will include a contribution which represents a global, time-dependent phase applied to $\cond$. We retain the freedom to choose this phase by including a variation in the parameter $\lambda_0(t)$
\beq
\lambda_0(t) = \lambda_0 + \dl_{0}(t).
\eeq
Substituting these expressions into Eq.~(\ref{TDGPE2}) and linearizing with respect to $\delta \Phi_0 \rt$, $\dl_{0}(t)$ and $P\rt$ gives the equation of motion for the fluctuation
\begin{align}
i \hbar \frac{\partial}{\partial t} \delta \Phi_0 \rt &= \hLGP \delta \Phi_0 \rt + \MGP \delta \Phi_0^* \rt \nonumber \\
&+ \bigl [P\rt -\dl_{0} (t) \bigr ] \cond,
\label{dphi0t}
\end{align}
where the quantities $\hLGP$ and $\MGP$ are defined similarly to $\hat{L} _0(\bfr)$ and $\hat{M} _0(\bfr)$ of Eqs.~(\ref{L0r}) and (\ref{M0r}) but with the orthogonal projector replaced by unity
\begin{align}
\hLGP &= \hsp(\bfr) -\lambda_0 + 2N_0U_0 |\cond|^2,\\
\MGP &= N_0U_0 \Phi_0^2 (\bfr). \label{M0}
\end{align}

Combining Eq.~(\ref{dphi0t}) with its complex conjugate and taking the Fourier transform, we obtain
\begin{align}
\left [ \hbar \omega - \cLGP_0(\bfr)\right ]\mathcal{D}\Phi_0\rw& \nonumber \\
- \bigl [ \bigr. P\rw-\dl_{0}(\omega)& \bigl. \bigr ] \mathcal{X}_{\theta_0}(\bfr) = 0,\label{dPhi0w_matrix}
\end{align}
where the condensate fluctuation spinor $\mathcal{D}\Phi_0\rw$ is
\beq
\mathcal{D}\Phi_0\rw = \bpm \delta\Phi_0\rw \\ \delta\Phi_0^*\rmw \epm,
\label{def_DPhi_0}
\eeq
and $\cLGP_0(\bfr)$ is defined analogously to $\mathcal{L}_0(\bfr)$
\beq
\cLGP_0(\bfr) = \begin{pmatrix} \hLGP & \MGP \\ -M_{0}^{\text{(GP)}*}(\bfr) & -\hLGP \end{pmatrix}.
\label{LGP}
\eeq
The Fourier transforms are defined by $\delta \Phi_0 \rt = \int_{-\infty}^{\infty} d\omega \, \delta \Phi_0 \rw e^{- i \omega t}$, and we have used the fact that the components of $\delta \Phi_0 \rt$ and $\delta \Phi_0^*\rt$ oscillating at the frequency $\omega$ are $\delta \Phi_0 \rw$ and $\delta \Phi_0^* (\bfr,-\omega)$ respectively.

We solve Eq.~(\ref{dPhi0w_matrix}) by expanding the fluctuation in the complete set of states provided by the eigenstates of $\cL_0(\bfr)$ given in Eq.~(\ref{LBdG_matrix})
\beq
\mathcal{D}\Phi_0\rw = \sum_{i} b_i(\omega)\mathcal{X}_i(\bfr),
\eeq
where the sum is over all states (i.e. with both positive and negative norms, zero and non-zero energy). The form of $\mathcal{D}\Phi_0\rw$ and the relation between positive and negative energy spinors of Eq.~(\ref{X_minus_i}) implies that $b_{-i}(\omega) = b_i^*(-\omega)$ [in the time-domain $b_{-i}(t) = b_i^*(t)$]. The expansion coefficients $\{ b_i(\omega) \}$ can be projected out of this expression using the orthonormality relation of Eq.~(\ref{orthosym}). Specifically, we have
\beq
b_i(\omega) = \chi_i\intdr \mathcal{X}_i^{\dagger}(\bfr)\sigma_3 \mathcal{D}\Phi_0\rw.
\label{project_b}
\eeq

It is convenient to separate the contribution from the two zero-energy condensate modes, and write the expansion in terms of the norm and phase states defined in Eq.~(\ref{def_Xnorm_Xphase}). Hence we write
\begin{align}
\mathcal{D}\Phi_0\rw &= b_{N_0}(\omega)\mathcal{X}_{N_0}(\bfr) + ib_{\theta_0}(\omega)\mathcal{X}_{\theta_0}(\bfr) \nonumber \\
&\phantom{=} + \sum_{i \neq \pm 0} b_i(\omega)\mathcal{X}_i(\bfr),
\label{qp_expansion}
\end{align}
where the sum is now over all non-condensate modes and $b_{N_0}(\omega)$ and $b_{\theta_0}(\omega)$ are respectively the Fourier transforms of the real and imaginary parts of $b_{+0}(t)$ (the coefficient of the positive norm, condensate mode in the time-domain). The first term in this expansion corresponds to changes in the norm of the condensate while the second corresponds to changes in phase. Conservation of the norm of the condensate wavefunction gives $b_{N_0}(\omega) = 0$ and we will choose the parameter $\dl_{0}(\omega)$ so that there is no phase mode, i.e. $b_{\theta_0}(\omega) = 0$ (see below).

The action of $\cLGP_0(\bfr)$ on the 
$\mathcal{X}_i(\bfr)$ spinors is easily found using the fact that these are eigenstates of $\cL_0(\bfr)$. For $\epsilon_i \neq 0$ we have
\beq
\cLGP_0(\bfr)\mathcal{X}_i(\bfr) = \epsilon_i \mathcal{X}_i(\bfr) + c_i\mathcal{X}_{\theta_0}(\bfr),
\label{LGP_X}
\eeq
where the coefficients $c_i$ are the time-independent versions of Eq.~(\ref{c_coeffs_t}) with $\Phi_2\rt \rightarrow \Phi_0(\bfr)$ \cite{Morgan00}
\beq
c_i = \intdr N_0 U_0 |\cond|^2 \bigl [ \ccond u_i(\bfr) + \cond v_i(\bfr) \bigr ]. \label{c_coeffs}
\eeq
For the phase and norm modes we have
\begin{gather}
\cLGP_0(\bfr)\mathcal{X}_{\theta_0}(\bfr) = 0,\\
\cLGP_0(\bfr)\mathcal{X}_{N_0}(\bfr) = 2N_0U_0|\cond|^2\mathcal{X}_{\theta_0}(\bfr).
\end{gather}
Equation~(\ref{c_coeffs}) shows that the $c_i$ coefficients are only non-zero for modes with the same symmetry as the condensate. In particular, they vanish in the homogeneous limit and are only non-zero for gases in anisotropic, harmonic traps for modes with positive z-parity and zero axial angular momentum ($m = 0$). If $c_i = 0$, Eq.~(\ref{LGP_X}) shows that the corresponding spinor $\mathcal{X}_i(\bfr)$ is an eigenstate of $\cLGP_0(\bfr)$ as well as of $\cL_0(\bfr)$.

Substituting the expansion of Eq.~(\ref{qp_expansion}) into Eq.~(\ref{dPhi0w_matrix}) and projecting out the coefficients using Eqs.~(\ref{project_b}) and (\ref{overlap_Xnorm_Xphase}) gives
\begin{gather}
b_p(\omega) = \frac{\chi_p P_{p0}(\omega)}{\hbar \omega-\epsilon_p},
\label{bp_GPE}\\
i\hbar \omega b_{\theta_0}(\omega) = P_{00}(\omega) - \dl_{0}(\omega)+ \sum_{i \neq \pm0} b_i (\omega) c_i,
\label{b0I}
\end{gather}
where the excitation matrix elements $P_{00}(\omega)$ and $P_{p0}(\omega)$ are defined by
\begin{align}
P_{00}(\omega) &= \intdr P \rw |\Phi_0|^2, \label{P_00}\\
P_{p0}(\omega) &= \intdr P \rw \left [ u_{p}^{*} \Phi_0 + v_{p}^{*} \Phi_0^* \right ],\label{P_p0}
\end{align}
($P_{00}(\omega)$ is just a special case of $P_{p0}(\omega)$ corresponding to $p=\pm0$). Eq.~(\ref{b0I}) shows that the choice
\beq
\dl_{0}(\omega) = P_{00}(\omega) + \sum_{i \neq \pm0} c_i b_i (\omega),
\label{dl0}
\eeq
gives $b_{\theta_0}(\omega) = 0$ and so removes global phase fluctuations of the condensate. The fact that the coefficients $c_i$ are only non-zero for modes with the same symmetry as the condensate, means that a limited subset of modes contribute to $\dl_{0}(\omega)$. In the homogeneous limit all these coefficients are zero and therefore the issue of condensate phase fluctuations does not arise if $P_{00}(\omega) = 0$. 

The coefficients $\{b_i(\omega)\}$ describe the contribution to the condensate density response from the various quasiparticle modes. The condensate density fluctuations $\delta n_c\rt = \delta (N_0 |\Phi_0 \rt|^2)$ can be written in terms of these quantities using
\beq
\delta n_c({\bf r},\omega) = N_0\!\!\sum_{i \neq \pm0} \! b_i(\omega)\bigl [\ccond u_i({\bf r}) + \cond v_i({\bf r}) \bigr ].
\label{dnc_GPE}
\eeq
This expression provides the link between theory and experiment because experiments generally measure condensate density fluctuations, whereas the theory gives expressions for the coefficients $b_i(\omega)$.

The coefficient $b_p(\omega)$ in Eq.~(\ref{bp_GPE}) diverges if the condensate is driven at the resonance frequency of the corresponding excitation, $\omega = \epsilon_p/\hbar$. This is, of course, because there is no damping in the theory at this level of approximation. Finite quantities can be obtained using the standard prescription of inserting a small imaginary part in the denominator via $\omega \rightarrow \omega + i \gamma$. This procedure is essential for numerical calculations at the next level of approximation, and can be justified from the finite experimental observation time, as we discuss in Appendix~\ref{app_exp_resolution}.

\subsection{Generalized GPE theory} \label{LINEAR_GGPE}

For a given experimental configuration (type of atom, trapping potential etc) the analysis given above has only one parameter, namely the condensate population $N_0$. It is therefore effectively a zero-temperature treatment, making no explicit mention of the non-condensate. In this section, we extend the analysis to finite temperature using the GGPE to describe the dynamic coupling between the condensate and non-condensate. The previous results acquire an implicit temperature dependence because we use basis quasiparticle wavefunctions $\mathcal{X}_i(\bfr)$ and energies $\epsilon_i$ corresponding to the relevant equilibrium condensate population, $N_0 = N_0(T)$. However, in the following discussion, we do not write the temperature dependence explicitly and simply denote the equilibrium condensate population by $N_0$. As discussed at the start of Sec.~\ref{sec_linear_response_theory}, the theory developed in this section contains the leading order corrections to the Bogoliubov theory in the small parameter of Eq.~(\ref{validity}).

The starting point for the higher order calculation is the GGPE
of Eq.~(\ref{TDGGPE}) with $\hsp\rt = \hsp(\bfr) + P\rt$ and UV renormalization as in Eq.~(\ref{rmt}). In the static case, the time-independent solutions $\Phi_2 \rt = \gcond$ satisfy
\begin{align}
 \left[ \hsp(\bfr) - \lambda_2 + (N_0 + \Delta N_0)U_0|\gcond|^2 \right] \gcond& \label{TIGGPE}\\
+ 2U_0\nt(\bfr)\gcond + U_0\mt^{\text{R}}(\bfr)\gccond - f(\bfr)& = 0, \nonumber
\end{align}
where $\lambda_2$ is a constant, again roughly equal to the chemical potential in an improved approximation. $\nt(\bfr)$, $\mt^{\text{R}}(\bfr)$ and $f(\bfr)$ are calculated from Eqs.~(\ref{nt_qp})-(\ref{N_i_kT}) and (\ref{rmt}) using static quasiparticle wavefunctions and energies obtained by solving Eq.~(\ref{LBdG_matrix}) with $N_0 = N_0(T)$.

As before, application of the external perturbation $P\rt$ leads to all quantities developing a small time-dependent oscillation around their static values. We therefore write
\beq
\left.
\begin{aligned}
\hsp\rt &= \hsp (\bfr) + P\rt, \\
\Phi_2 \rt &= \gcond + \delta \Phi_2 \rt,\\
\lambda_2(t) &= \lambda_2 + \dl_{2}(t),\\
N_0(t) &= N_0 + \delta N_0(t),\\
\nt \rrt &= \nt\rr + \delta\nt\rrt,\\
\mt^{\text{R}} \rrt &= \mt^{\text{R}}\rr + \delta\mt^{\text{R}}\rrt,\\
f \rt &= f(\bfr) + \delta f \rt.
\end{aligned}
\right \}
\label{GGPE_linearizations}
\eeq
We neglect here any dynamic change in the variance of the condensate population $\Delta N_0$, although changes in the mean are included via $\delta N_0(t)$. Substituting these expressions into Eq.~(\ref{TDGGPE}) and linearizing with respect to all small quantities, leads to an equation of motion for the condensate fluctuation $\delta \Phi_2 \rt$. As before we take the Fourier transform of this and combine it with its complex conjugate in the two-component form $\mathcal{D}\Phi_2\rw$ defined in the same manner as Eq.~(\ref{def_DPhi_0}).

Linearizing with respect to the quantities in Eq.~(\ref{GGPE_linearizations}) is exact in the limit $P \rt \rightarrow 0$. However, we now perform a second linearization which is governed by the small parameter of Eq.~(\ref{validity})
\begin{align}
\gcond &= \cond + \Delta \Phi_{20} (\bfr),\label{dPhi20}\\
\lambda_2 &= \lambda_0 + \Delta \lambda_{20}.\label{dlambda20}
\end{align}
$\Delta \Phi_{20} (\bfr)$ and $\Delta \lambda_{20}$ are static quantities which arise from the difference in the condensate shape and energy between the generalized and ordinary GPE descriptions. In all terms which were not present in the GPE theory of the previous section, we make the leading order replacement $\gcond \rightarrow \cond$ and $\lambda_2 \rightarrow \lambda_0$, while in terms which were present in the earlier treatment we keep $\Delta \Phi_{20} (\bfr)$ and $\Delta \lambda_{20}$ to linear order. We note that because these corrections are independent of $P\rt$ we must retain quantities such as $\Delta \Phi_{20} (\bfr)\delta \Phi_2 \rt$. These are not quadratic in the perturbation, but instead are linear in the perturbation and also linear in the small parameter of the theory.

Carrying out this second linearization, leads to the following equation for the condensate fluctuations
\beq
\begin{split}
\left [ \hbar \omega \right.&-\left.\cLGP_0(\bfr) \right ] \mathcal{D}\Phi_2\rw - \left[ P\rw-\dl_{2}^{r}(\omega) \right] \mathcal{X}_{\theta_0}(\bfr)\\
&= \left[ P\rw-\dl_{2}^{r}(\omega) \right] \begin{pmatrix} \Delta \Phi_{20} (\bfr) \\ -\Delta \Phi_{20}^{*} (\bfr) \end{pmatrix} -i\dl_{2}^{i}(\omega)\mathcal{X}_{N_0}(\bfr) \\
&\phantom{=}+ \begin{pmatrix} \Delta L_{20}^{\text{(GP)}}(\bfr) & \Delta M_{20}^{\text{(GP)}}(\bfr) \\ -\Delta M_{20}^{\text{(GP)}*} (\bfr) & -\Delta L_{20}^{\text{(GP)}}(\bfr) \end{pmatrix}\mathcal{D}\Phi_2\rw \\
&\phantom{=}+ \begin{pmatrix} 2U_0\delta\nt\rw & U_0\delta\mt^{\text{R}}\rw \\ - U_0\delta\mt^{\text{R}*}\rmw & -2U_0\delta\nt\rw \end{pmatrix}\mathcal{X}_{N_0}(\bfr)\\
&\phantom{=}- \begin{pmatrix}\delta f \rw \\ -\delta f^* \rmw \end{pmatrix} + \delta N_0(\omega)U_0 |\cond|^2\mathcal{X}_{\theta_0}(\bfr).
\end{split}
\label{dPhi2w_matrix} 
\eeq
Here $\Delta L_{20}^{\text{(GP)}}(\bfr)$ and $\Delta M_{20}^{\text{(GP)}}(\bfr)$ are 
defined by
\begin{align}
\Delta L_{20}^{\text{(GP)}}(\bfr) &= 2U_0\nt(\bfr) - \Delta \lambda_{20}+ 2\Delta N_0U_0|\cond|^2  \\
&+ 2N_0U_0 \bigl [ \cond \Delta\Phi_{20}^{*}(\bfr) +  \ccond\Delta\Phi_{20}(\bfr) \bigr ],\nonumber
\end{align}
\begin{align}
\Delta M_{20}^{\text{(GP)}}(\bfr) = U_0 \mt^{\text{R}}(\bfr) + \Delta N_0U_0\Phi_0^2(\bfr)& \nonumber \\
+ 2N_0U_0 \cond \Delta\Phi_{20}(\bfr)&,
\end{align}
and $\dl_{2}^{r}(\omega)$ and $\dl_{2}^{i}(\omega)$ are respectively the Fourier transforms of the real and imaginary parts of $\dl_2(t)$. In fact, the imaginary part of $\dl_2(t)$ makes no contribution to the evolution of a mode with finite energy because it appears multiplied by the condensate norm mode which is orthogonal to all such states.

We solve Eq.~(\ref{dPhi2w_matrix}) as before by expanding the fluctuation in the static quasiparticle basis of Eq.~(\ref{LBdG_matrix})
\begin{align}
\mathcal{D}\Phi_2\rw &= b_{N_0}(\omega)\mathcal{X}_{N_0}(\bfr) + ib_{\theta_0}(\omega)\mathcal{X}_{\theta_0}(\bfr) \nonumber \\
&\phantom{=} + \sum_{i \neq \pm 0} b_i(\omega)\mathcal{X}_i(\bfr).
\label{qp_expansion2}
\end{align}
Unlike the previous expansion of Eq.~(\ref{qp_expansion}), however, the coefficient $b_{N_0}(\omega)$ is not zero, even though the norm of $\Phi_2 \rt$ is conserved by the GGPE. The reason is that the condensate wavefunction appearing in the condensate modes of the expansion is $\cond$ and not $\gcond$. In fact, conservation of the norm of $\Phi_2 \rt$ gives
\beq
b_{N_0}(\omega) = -\frac{1}{2}\sum_{i\neq\pm0} \! b_i(\omega)\!\intdr \bigl ( \Delta \Phi_{20}^{*}u_i + \Delta \Phi_{20}v_i \bigr ).
\label{b0R} 
\eeq
In this equation we have used the fact that $b_{N_0}(\omega)$ is zero at leading order and also that the relative phase of $\gcond$ and $\cond$ can be chosen so that the phase mode gives no contribution. As in the GPE calculation, we can always choose $\dl_2^{r}(\omega)$ so that $b_{\theta_0}(\omega) = 0$. At leading order this requires choosing $\dl_2^{r}(\omega) = \dl_0(\omega)$ as given by Eq.~(\ref{dl0}) and this expression is sufficient for this calculation. We will assume from here on that this choice for $\dl_2^r(\omega)$ has been made.

Substituting Eq.~(\ref{qp_expansion2}) into Eq.~(\ref{dPhi2w_matrix}) and projecting out the coefficient of mode `p' $\neq \pm 0$ gives
\begin{multline}
\chi_p (\hbar \omega - \epsilon_p)b_p(\omega) - P_{p0}(\omega) = \Delta P_{p0}^{(S)}(\omega) + \Delta P_{p0}^{(D)}(\omega) \\
+ \Delta E_{p}^{(S)}(\omega) + \Delta E_{p}^{(D)}(\omega),
\label{bp_GGPE}
\end{multline}
where the various quantities on the right-hand-side are defined and discussed below. These new terms correspond to changes in the excitation matrix element and energy of the quasiparticle mode `p' arising from the presence of the non-condensate. The changes are of two types, static and dynamic, denoted by the superscripts $(S)$ and $(D)$ respectively, and correspond to the different roles of the thermal cloud. The static term $\Delta E_{p}^{(S)}(\omega)$ comes from interactions between a condensate fluctuation and the equilibrium non-condensate mean-fields $\nt(\bfr)$, $\mt^{\text{R}}(\bfr)$ and $f(\bfr)$. $\Delta P_{p0}^{(S)}(\omega)$ describes the effect of changes in the static condensate shape [$\Phi_0(\bfr) \rightarrow \Phi_2(\bfr)$] on the excitation matrix element $P_{p0}(\omega)$. Dynamic terms describe the effect of the coupling between condensate and non-condensate fluctuations.

Non-condensate dynamics can occur via two mechanisms; either the non-condensate is driven directly by the perturbation or it is driven indirectly via the condensate. The second possibility is the one usually considered in analytical discussions of condensate excitations. This mechanism gives rise to the dynamic term $\Delta E_{p}^{(D)}(\omega)$ describing the generation of non-condensate fluctuations by the condensate and their subsequent back action which damps and shifts the condensate excitations. The inclusion of this contribution is required to cancel infra-red divergences in the static terms and to obtain a gapless excitation spectrum consistent with the Hugenholtz-Pines theorem \cite{Hugenholtz59}. This is demonstrated explicitly in \cite{Giorgini00,Fedichev98,Morgan00} and is discussed further in Sec.~\ref{sec_homogeneous}.

However, the non-condensate can also be excited directly by the external perturbation. The resulting fluctuations can then couple to the condensate and act, in effect, as an additional external perturbation. This process therefore changes the excitation matrix element $P_{p0}(\omega)$ and is described by the term $\Delta P_{p0}^{(D)}(\omega)$. Direct excitation of the non-condensate is generally neglected in analytical discussions but turns out to be of crucial importance in explaining the results from the 1997 JILA experiment, as demonstrated recently \cite{Jin97,Morgan03a}. It is also important to include this effect for consistency with the dipole mode, see Sec.~\ref{sec_dipole_mode} and \cite{Morgan03c}.

The static and dynamic shifts can be written as an expansion over the various excited quasiparticle modes as
\begin{align}
\Delta E_{p}^{(S)}(\omega) &= \sum_{q\neq \pm 0} \Delta E_{pq}^{(S)}b_q(\omega),\label{DEs}\\
\Delta E_{p}^{(D)}(\omega) &= \sum_{q\neq \pm 0} \Delta E_{pq}^{(D)}(\omega)b_q(\omega),\label{DEd}
\end{align}
where the quantities $\Delta E_{pq}^{(S)}$ are independent of frequency while the quantities $\Delta E_{pq}^{(D)}(\omega)$ have a non-trivial frequency dependence. Equation~(\ref{bp_GGPE}) can therefore be written as a matrix eigenvalue equation for non-condensate modes ($p,q \neq \pm 0$)
\begin{subequations}
\begin{gather}
\sum_{q \neq \pm 0}\mathcal{H}_{pq}(\omega)b_q(\omega) = \mathcal{P}_{p0}(\omega),\\
\mathcal{H}_{pq}(\omega) = \chi_p \left ( \hbar\omega-\epsilon_p \right )\delta_{pq} - \Delta E_{pq}^{(S)} - \Delta E_{pq}^{(D)}(\omega),\\[5pt]
\mathcal{P}_{p0}(\omega) = P_{p0}(\omega) + \Delta P_{p0}^{(S)}(\omega) + \Delta P_{p0}^{(D)}(\omega),
\end{gather}
\label{bp_GGPE_matrix}
\end{subequations}
where $\delta_{pq}$ is the Kronecker delta.

This equation can be solved by inversion of the matrix $\mathcal{H}(\omega)$. We simplify the calculation considerably, however, by assuming that the perturbation $P\rt$ has a symmetry and frequency width such that only one quasiparticle mode is significantly excited. This is appropriate for experiments which measure the excitation energies of the system as a function of temperature. In addition, if we are only interested in frequencies near a particular unperturbed resonance $\omega \approx \epsilon_p/\hbar$, we can use a rotating-wave-approximation and neglect the contribution from the mode with energy $-\epsilon_p$. In this case, we can assume that only a single coefficient $b_p(\omega)$ is non-zero and Eq.~(\ref{bp_GGPE_matrix}) can be solved straightforwardly. Introducing the experimental resolution via $\omega \rightarrow \omega+i\gamma$ as discussed in Appendix~\ref{app_exp_resolution} we find that the Fourier transform of the response coefficient can be written as
\beq
b_p(\omega) = \chi_p P_{p0}(\omega)\mathcal{R}_p(\omega+i\gamma),
\label{bp_GGPE_soln}
\eeq
where the generalized response function $\mathcal{R}_p(\omega)$ is defined by
\beq
\mathcal{R}_p(\omega) = \left [ 1 + \frac{\Delta P_{p0}^{(S)}(\omega) + \Delta P_{p0}^{(D)}(\omega)}{P_{p0}(\omega)} \right ]{\cal G}_p(\omega),
\label{R}
\eeq
and the resolvent $\mathcal{G}_p(\omega)$ can be written in terms of a frequency-dependent self-energy as
\begin{align}
\mathcal{G}_p(\omega) &= \frac{1}{\hbar \omega - \epsilon_p - \Sigma_p(\omega)},
\label{G}\\
\Sigma_p(\omega) &= \chi_p \left [ \Delta E_{pp}^{(S)} + \Delta E_{pp}^{(D)}(\omega) \right ].
\label{sigmapw}
\end{align}

$\mathcal{G}_p(\omega)$ describes the effect of the static and dynamic coupling between the condensate and non-condensate in the limit that the external perturbation only excites the condensate. The additional factor in $\mathcal{R}_p(\omega)$ accounts for the change in the excitation matrix element when non-condensate fluctuations are excited directly by the external perturbation. The self-energy $\Sigma_p(\omega+i\gamma)$ is a complex function, the real part describing a shift in the energy of the excitation and the imaginary part giving its damping.

If $\Sigma_p(\omega+i\gamma)$ and $\Delta P_{p0}^{(D)}(\omega+i\gamma)$ are roughly independent of frequency, the energy shift can be calculated straightforwardly from the poles of $\mathcal{G}_p(\omega+i\gamma)$ by finding the solutions to
\beq
E_p = \hbar \omega_p = \mbox{Real}\bigl [ \epsilon_p+\Sigma_p(\omega_p+i\gamma) \bigr ].
\label{E_p}
\eeq
The corresponding decay rate is then given by
\beq
\Gamma_p = -\mbox{Imag}\bigl [\Sigma_p(\omega_p+i\gamma) \bigr ]/\hbar.
\label{Gamma_p}
\eeq
This situation arises when an excitation couples to a continuum of decay channels, as in the homogeneous limit, and the resolvents $\mathcal{G}_p$ and $\mathcal{R}_p$ are Lorentzians. For a finite system, however, $\Sigma_p(\omega+i\gamma)$ depends on frequency, and neither $\mathcal{G}_p$ nor $\mathcal{R}_p$ are perfect Lorentzians. In this case the line shape depends on the details of the system, in particular the availability of decay channels, the coupling strength to each of them, and the broadening of the levels involved. We can still extract energies and decay rates, however, by fitting $b_p(\omega)$ to a complex Lorentzian (the experimental resolution $\gamma$ should be subtracted from the resulting decay rate). An alternative is to mimic the experimental procedure exactly and Fourier transform appropriate moments of the condensate density response (Eq.~(\ref{dnc_GGPE}) below) to the time-domain and fit a decaying sinusoid.

As before, the coefficients $\{b_p(\omega)\}$ can be used to construct the condensate density fluctuations, which are the experimentally relevant quantity. At this order, however, we must include both the dynamic fluctuations in the condensate number $\delta N_0(\omega)$ and also the non-zero coefficient $b_{N_0}(\omega)$. Equation~(\ref{dnc_GPE}) is therefore replaced by
\begin{align}
\delta n_c({\bf r},\omega) &= \bigl [\delta N_0(\omega) + 2N_0b_{N_0}(\omega)\bigr ]|\cond|^2 \label{dnc_GGPE} \\
&+ N_0 \!\!\sum_{i \neq \pm 0} \! b_i(\omega)\bigl [\ccond u_i({\bf r}) + \cond v_i({\bf r}) \bigr ],\nonumber
\end{align}
where $\delta N_0(\omega)$ is defined by $\delta N_0(\omega) = -\intdr \delta \nt\rw$ and is given explicitly in Eq.~(\ref{app_dN0}). We can also calculate the spatial dependence of the non-condensate density fluctuations $\delta\nt\rw$ using the explicit expression given in Eq.~(\ref{app_dnrrw}). This can then be used to investigate the issue of the relative phase of condensate and non-condensate oscillations \cite{Bijlsma99,Khawaja00,Morgan03c}.

We now define the various quantities appearing in Eqs.~(\ref{bp_GGPE})-(\ref{DEd}) and discuss their physical interpretation. Details concerning the derivation of the expressions are given in Appendix~\ref{app_energy_shifts}.

\subsubsection{Definition of static shifts}

We define static shifts as those terms which do not involve any dynamic non-condensate fluctuations. The only frequency dependence in these terms therefore arises from the expansion coefficients $\{b_i(\omega)\}$.

The energy shift $\Delta E_{pq}^{(S)}$ of Eq.~(\ref{DEs}) can be written as
\begin{multline}
\Delta E_{pq}^{(S)} = \Delta E_{4}(p,q) + \Delta E_{\lambda}(p,q) + \Delta E_{\text{shape}}(p,q)\\
+ \Delta E_{\Delta N_0}(p,q) + \Delta E_{f}^{(S)}(p,q),
\end{multline}
where the various contributions are defined by
\begin{align}
\Delta E_{4}(p,q) = \intdr \Bigl \{ \Bigr. 2&U_0\nt \left [ u_p^*u_q +  v_p^*v_q \right ] \nonumber\\
+ &U_0\mt^{\text{R}} u_p^*v_q + U_0\mt^{\text{R}*}v_p^*u_q \Bigl. \Bigr \},\label{de4}
\end{align}
\beq
\Delta E_{\lambda}(p,q) = -\Delta \lambda_{20} \intdr \left [ u_p^*u_q +  v_p^*v_q \right ],
\label{delambda}
\eeq
\begin{align}
\Delta E_{\text{shape}}(p,q) &= \nonumber \\
2N_0 U_0 \!\! &\intdr \Bigl\{\Bigr. [ \Phi_0 \Delta \Phi_{20}^{*} + \Phi_0^* \Delta\Phi_{20} ] [ u_p^*u_q +  v_p^*v_q ] \nonumber \\
 & \qquad + [ \Phi_0 \Delta\Phi_{20} u_p^*v_q + \Phi_0^* \Delta\Phi_{20}^{*} v_p^*u_q ] \Bigl.\Bigr \} \nonumber \\
-&\intdr \Bigl\{\Bigr. c_p^*[ \Delta\Phi_{20}^{*} u_q + \Delta\Phi_{20} v_q ] \nonumber \\
 & \qquad +c_q[ \Delta\Phi_{20} u_p^* + \Delta\Phi_{20}^* v_p^* ]\Bigl.\Bigr \}, \label{deshape}
\end{align}
\begin{align}
\Delta E_{\Delta N_0}(p,q) = \Delta N_0U_0 \!\! \intdr \Bigl \{ \Bigr. 2|&\Phi_0|^2\left [ u_p^*u_q +  v_p^*v_q \right ] \nonumber \\ 
+ &\Phi_0^2 u_p^* v_q + \Phi_0^{*2} v_p^* u_q \Bigl. \Bigr \},
\label{destat}
\end{align}
\begin{align}
\Delta E_{f}^{(S)}(p,q) =& \nonumber\\
-\sum_{i>0}& \frac{\left ( A_{qi0}+B2_{q0i} \right )}{2\sqrt{N_0}} \left [ N_iW_{pi}^* + (N_i + 1)W_{pi}^* \right ] \nonumber \\
-\sum_{i>0}&\frac{\left ( B1_{qi0} + B2_{qi0} \right )}{2\sqrt{N_0}}\left [ N_i U_{pi} + (N_i + 1)V_{pi} \right ].\label{defs}
\end{align}
The various quantities in $\Delta E_{f}^{(S)}(p,q)$ are defined by
\begin{align}
U_{ij} &= \intdr u_i^*(\bfr)u_j(\bfr), \label{Uij}\\
V_{ij} &= \intdr v_i^*(\bfr)v_j(\bfr), \label{Vij} \\
W_{ij} &= \intdr u_i(\bfr)v_j(\bfr), \label{Wij}
\end{align}
while $A_{qi0}$, $B1_{qi0}$, $B2_{qi0}$ and $B2_{q0i}$ are special cases of the coefficients $A_{qij}$, $B1_{qij}$ and $B2_{qij}$ defined later in Eqs.~(\ref{A})-(\ref{B2}) [the index `0' refers to the condensate mode with positive norm of Eq.~(\ref{X0plus})].

The quantities $\Delta\lambda_{20}$ and $\Delta\Phi_{20}(\bfr)$ needed to evaluate $\Delta E_{\lambda}(p,q)$ and $\Delta E_{\text{shape}}(p,q)$ are calculated in Appendix~\ref{app_condensate_statics}. Briefly, the static solution to the GGPE of Eq.~(\ref{TIGGPE}) is obtained by linearizing around the static solution to the ordinary GPE using Eqs.~(\ref{dPhi20}) and (\ref{dlambda20}). The resulting equation is solved using an expansion in the quasiparticle basis of Eq.~(\ref{LBdG_matrix}). This linearization is necessary for consistency with the linearized treatment of condensate shape and energy effects in Eq.~(\ref{dPhi2w_matrix}) and also for a systematic treatment to a given order in the small parameter of the theory.

The static change in the excitation matrix element $\Delta P_{p0}^{(S)}(\omega)$ is defined by
\beq
\Delta P_{p0}^{(S)}(\omega) = \intdr \bigl [ P\rw - P_{00}(\omega)\bigr] \bigl [ u_p^* \Delta\Phi_{20} + v_p^* \Delta\Phi_{20}^* \bigr ].
\label{DP_p0_s}
\eeq
The contribution from $P_{00}(\omega)$ arises from the choice of $\delta \lambda_2^r(\omega) = \delta \lambda_0(\omega)$ given in Eq.~(\ref{dl0}) which removes condensate phase fluctuations to leading order. The rest of the expression is a straightforward modification of $P_{p0}(\omega)$ as given in Eq.~(\ref{P_p0}) accounting for the change in shape of the condensate.

\subsubsection{Interpretation of static shifts}

We now discuss the physical interpretation of the various terms appearing in the static shifts. $\Delta E_{4}(p,q)$ is the energy shift arising from the interaction between the condensate fluctuation and the static mean-fields of the non-condensate $\nt (\bfr)$ and $\mt^{\text{R}}(\bfr)$. In an alternative approach to the theory of the Bose gas, based on perturbation theory with the many-body Hamiltonian, this contribution arises from the use of first order perturbation theory on products of four non-condensate operators (hence the choice of notation) \cite{Morgan00}. $\Delta E_{f}^{(S)}(p,q)$ arises from the definition of the non-condensate as orthogonal to the condensate while $\Delta E_{\Delta N_0}(p,q)$ expresses the effect of the variance in the condensate population on the nonlinear term in the GPE.

$\Delta E_{\lambda}(p,q)$ and $\Delta E_{\text{shape}}(p,q)$ both arise from changes in the properties of the equilibrium condensate wavefunction due to interactions with the static non-condensate mean-fields. The condensate eigenvalue $\lambda$ is the zero of energy for excitations so a change in this gives rise to $\Delta E_{\lambda}(p,q)$. The first two lines of $\Delta E_{\text{shape}}(p,q)$ have a straightforward interpretation, describing the change in the condensate mean-field potential when the condensate shape changes. The final two lines come from a more subtle effect. In the current formulation these arise from the non-zero condensate norm coefficient $b_{N_0}(\omega)$ given in Eq.~(\ref{b0R}) and also from the choice of $\delta\lambda_2^r (\omega)$ which removes phase fluctuations (see Appendix~\ref{app_static_shifts} for the details). In \cite{Morgan00}, the same contribution is obtained from the fact that quasiparticle excitations are defined to be orthogonal to the condensate. A change in the shape of the condensate therefore leads to a change in the definition of the orthogonal subspace and this in turn affects the energy of the excitations.

\subsubsection{Definition of dynamic shifts}

We define dynamic terms as those which involve fluctuations in the non-condensate. These terms have a non-trivial frequency dependence over and above that contained in the expansion coefficients $\{b_i(\omega)\}$, as indicated in Eq.~(\ref{DEd}). 

Expressions for the energy shift $\Delta E_{pq}^{(D)}(\omega)$ and the change in the excitation matrix element $\Delta P_{p0}^{(D)}(\omega)$ are derived in Appendix~\ref{app_dynamic_shifts}. Briefly, the derivation consists of writing $\nt \rrt$ and $\mt \rrt$ in terms of the time-dependent Bogoliubov functions $\mathcal{X}_i\rt$ and evolving these according to Eq.~(\ref{uvt}). Linearizing the changes in these quantities around their static values and solving the resulting equations in terms of the static Bogoliubov basis, leads eventually to
expressions for $\delta\nt\rrw$ and $\delta\mt\rrw$ from which $\Delta E_{pq}^{(D)}(\omega)$ and $\Delta P_{p0}^{(D)}(\omega)$ follow straightforwardly.

The result for $\Delta E_{pq}^{(D)}(\omega)$ is
\begin{align}
\Delta E_{pq}^{(D)}(\omega) &= \sum_{ij > 0} -\frac{Y_{pij}^{(A)*}Y_{qij}^{(A)}}{2 \,( \hbar \omega + \epsilon_{i} + \epsilon_{j} )} \left [ 1 + N_{i} + N_{j} \right ] \nonumber \\
&+ \sum_{ij > 0} \frac{Y_{pij}^{(B1)*}Y_{qij}^{(B1)}}{2 \,( \hbar \omega -\epsilon_{i} - \epsilon_{j})} \left [1 + N_{i} + N_{j} \right ]\nonumber\\
&+ \sum_{ij > 0} \frac{Y_{pij}^{(B2)*}Y_{qij}^{(B2)}}{\hbar \omega -\epsilon_{i}+ \epsilon_{j}} \left [N_{j} - N_{i} \right ]\nonumber\\
&+ \Delta E_{0}(p,q) + \Delta E^{\text{R}}(p,q),
\label{ded_pq}
\end{align}
where $\Delta E_{0}(p,q)$ is defined by
\begin{align}
\Delta E_{0}(p,q) &= \nonumber \\
 -\frac{1}{\sqrt{N_0}} &\sum_{i>0} \Bigl [ Y_{p0i}^{(B2)*}W_{iq} N_i + Y_{pi0}^{(A)*}W_{iq}(N_i+1) \Bigr ] \nonumber \\
 -\frac{1}{\sqrt{N_0}} &\sum_{i>0} \Bigl [ Y_{pi0}^{(B2)*}U_{iq} N_i + Y_{pi0}^{(B1)*}V_{iq}(N_i+1)\Bigr ],
\label{de0}
\end{align}
and $\Delta E^{\text{R}}(p,q)$ is the UV renormalization
\beq
\Delta E^{\text{R}}(p,q) = 2N_0\Delta U_0\intdr |\Phi_0|^2 \left(u_p^*u_q + v_p^*v_q \right),
\label{deR}
\eeq
with $\Delta U_0$ as in Eq.~(\ref{DeltaU0}).

The coefficients $Y_{pij}^{(A)}$, $Y_{pij}^{(B1)}$ and $Y_{pij}^{(B2)}$ are defined by
\begin{subequations}
\begin{align}
Y_{pij}^{(A)} &= A_{pij} - \frac{c_p}{\sqrt{N_0}}J_{ij} - \frac{c_i}{\sqrt{N_0}}J_{pj} - \frac{c_j}{\sqrt{N_0}}J_{pi}, \label{YA} \\
Y_{pij}^{(B1)} &= B1_{pij} - \frac{c_p}{\sqrt{N_0}}J_{ij}^* - \frac{c_i^*}{\sqrt{N_0}}I_{pj}^* - \frac{c_j^*}{\sqrt{N_0}}I_{pi}^*, \label{YB1}\\
Y_{pij}^{(B2)} &= B2_{pij} - \frac{c_p}{\sqrt{N_0}}I_{ij} - \frac{c_i^*}{\sqrt{N_0}}J_{pj} - \frac{c_j}{\sqrt{N_0}}I_{pi}^*, \label{YB2}
\end{align}
\label{Ys}
\end{subequations}
where the coefficients $c_i$ are given in Eq.~(\ref{c_coeffs}) and $A_{pij}$, $B1_{pij}$, $B2_{pij}$, $I_{ij}$ and $J_{ij}$ are defined by \cite{gapless_coefficients}
\begin{align}
A_{pij} = 2\sqrt{N_{0}} U_{0} \! \intdr &u_{p} \bigl [ \Phi_0^* \left( u_{i}v_{j} + v_{i}u_{j} \right ) + \Phi_0 v_{i}v_{j} \bigr ] \nonumber \\
+ &v_{p} \bigl [ \Phi_0 \left( u_{i}v_{j} + v_{i}u_{j} \right) + \Phi_0^* u_{i}u_{j} \bigr ],\label{A}
\end{align}
\begin{align}
B1_{pij} = 2\sqrt{N_{0}} U_{0} \! \intdr & u_{p} \bigl [ \Phi_0^{*} \left( u_{i}v_{j} + v_{i}u_{j} \right)^{*}  + \Phi_0 u_{i}^{*}u_{j}^{*} \bigr ] \nonumber \\
+ &v_{p} \bigl [ \Phi_0 \left( u_{i}v_{j} + v_{i}u_{j} \right)^{*} + \Phi_0^{*} v_{i}^{*}v_{j}^{*} \bigr ],
\label{B1}
\end{align}
\begin{align}
B2_{pij} = 2\sqrt{N_{0}} U_{0} \! \intdr & u_{p} \bigl [ \Phi_0^* \left( u_{i}^{*}u_{j} + v_{i}^{*}v_{j} \right) + \Phi_0 u_{i}^{*}v_{j} \bigr ] \nonumber \\
+ &v_{p} \bigl [ \Phi_0 \left( u_{i}^{*}u_{j} + v_{i}^{*}v_{j} \right) + \Phi_0^* v_{i}^{*}u_{j} \bigr ],
\label{B2}
\end{align}
\begin{align}
I_{ij} &= \intdr \left[ u_i^* u_j + v_i^* v_j \right] = U_{ij}+V_{ij}, \label{Iij}\\
J_{ij} &= \intdr \left[ u_i v_j + v_i u_j \right] = 2W_{ij}.
\label{Jij}
\end{align}

The term $\Delta E_{0}(p,q)$ in Eqs.~(\ref{ded_pq}) and \ref{de0}), is the explicit contribution to $\Delta E_{pq}^{(D)}(\omega)$ from the condensate modes (note that the summations in Eq.~(\ref{ded_pq}) exclude the state $+0$). These modes appear in the description of $\delta\nt\rrw$ and $\delta\mt\rrw$ because the time-dependent Bogoliubov functions $\mathcal{X}_i\rt$ are orthogonal to the time-dependent condensate wavefunction (see Appendix~\ref{app_dynamic_shifts}).

The dynamic change in the excitation matrix element $\Delta P_{p0}^{(D)}(\omega)$ has a form similar to $\Delta E_{pq}^{(D)}(\omega)$
\begin{align}
\Delta P_{p0}^{(D)}(\omega) &= \sum_{ij>0} -\frac{Y_{pij}^{(A)*} P_{ij}^{(B)*}(-\omega)}{2 \sqrt{N_0}\,( \hbar \omega + \epsilon_{i} + \epsilon_{j} )} \left [ 1 + N_{i} + N_{j} \right ] \nonumber\\
&+ \sum_{ij>0} \frac{Y_{pij}^{(B1)*}P_{ij}^{(B)}(\omega)}{2 \sqrt{N_0} \,( \hbar \omega -\epsilon_{i} - \epsilon_{j})} \left [1 + N_{i} + N_{j} \right ]\nonumber
\\
&+ \sum_{ij>0} \frac{Y_{pij}^{(B2)*}P_{ij}^{(L)}(\omega)}{\sqrt{N_0}(\hbar \omega -\epsilon_{i}+ \epsilon_{j})} \left [N_{j} - N_{i} \right ],\label{dPp}
\end{align}
where we have defined Beliaev and Landau perturbation matrix elements by
\begin{align}
P_{ij}^{(B)}(\omega) &= \intdr \bigl [ P\rw - P_{00}(\omega) \bigr ] \bigl [ u_{i}^{*} v_{j}^{*} + v_{i}^{*} u_{j}^{*} \bigr ], \label{PBij}\\
P_{ij}^{(L)}(\omega) &= \intdr \bigl [ P\rw - P_{00}(\omega) \bigr ] \bigl [ u_{i}^{*} u_{j} + v_{i}^{*} v_{j} \bigr ],\label{PLij}
\end{align}
with $P_{00}(\omega)$ as in Eq.~(\ref{P_00}). To the best of our knowledge Eq.~(\ref{dPp}) has not been obtained previously in the literature.

To calculate the condensate density fluctuations in Eq.~(\ref{dnc_GGPE}) we also need an expression for $\delta N_0(\omega)$. This has the same structure as $\Delta E_{p}^{(D)}(\omega)$ and $\Delta P_{p0}^{(D)}(\omega)$ and is written explicitly in Eq.~(\ref{app_dN0}).

\subsubsection{Interpretation of dynamic terms}

The self-energy $\Delta E_{pq}^{(D)}(\omega)$ describes the effect of the dynamic coupling between the condensate and non-condensate in the collisionless limit. In this approximation, the non-condensate is modelled as a weakly-interacting gas of quasiparticles and its dynamics therefore corresponds to rearrangements of existing quasiparticles among the available levels (Landau processes) or the creation or annihilation of quasiparticles (Beliaev processes). More specifically, the first line of Eq.~(\ref{ded_pq}) corresponds to the collision and annihilation of three quasiparticles, while the second line describes the spontaneous decay of a single quasiparticle into two others. The third line corresponds to Landau processes in which two quasiparticles collide and coalesce to form a third. The matrix elements $Y_{pij}^{(A)}$, $Y_{pij}^{(B1)}$ and $Y_{pij}^{(B2)}$ describe the coupling strengths for these processes and contain the relevant selection rules (such as angular momentum conservation and parity). If the quasiparticle modes are thermally populated these processes are enhanced by Bose stimulation of the various scattering events involved. The dynamic shift thus corresponds physically to the introduction of many-body T-matrix effects into condensate-non-condensate collisions [the many-body T-matrix is introduced into condensate-condensate collisions by the static anomalous average $\mt^{\text{R}}(\bfr)$] \cite{Morgan00,Rusch99,Proukakis98b,Stoof93,Bijlsma97,Shi98}.

Landau and Beliaev processes can occur in a virtual sense or in a real sense when one of the energy denominators vanishes. To obtain finite quantities in calculations, it is therefore essential to include a small imaginary part in the frequency via $\omega \rightarrow \omega + i \gamma$, as in Eq.~(\ref{bp_GGPE_soln}). This is justified by the finite experimental resolution as shown in Appendix~\ref{app_exp_resolution}. $\Delta E_{pq}^{(D)}(\omega + i \gamma)$ is then a complex quantity, the real part describing the shift in the energy of a condensate fluctuation and the imaginary part describing its damping. 

We now consider the structure of $\Delta E_{p}^{(D)}(\omega)$ in more detail. The numerators in Eq.~(\ref{ded_pq}) contain a product of two of the $Y$ matrix elements. The various contributions to these terms, given in Eq.~(\ref{Ys}), have different physical origins depending on whether they appear with the label `$pij$' or `$qij$'. For example, the term involving $c_p$ in the $Y_{pij}$ factors comes from condensate number fluctuations, i.e. from $\delta N_0(\omega)$ in Eq.~(\ref{dPhi2w_matrix}). The corresponding term involving $c_q$ in the $Y_{qij}$'s instead comes from condensate phase fluctuations as shown in Appendix~\ref{app_dynamic_shifts}. There is thus an elegant symmetry in the way the theory deals with condensate phase and number fluctuations.

Similarly, the contribution from the terms involving $c_i$ and $c_j$ in the $Y_{pij}$ factors comes from the dynamic part of $\delta f\rw$ in Eq.~(\ref{dPhi2w_matrix}), while the corresponding terms in the $Y_{qij}$ factors come from changes in the orthogonal projectors $\hat{Q}$ in the equation of motion for the excited states. This demonstrates the connection between $f\rt$ and the issue of orthogonality between the condensate and non-condensate. The matrix elements $A_{qij}$ etc in the $Y_{qij}$'s describe the driving of the non-condensate by a condensate fluctuation in mode `q', while the corresponding terms in the $Y_{pij}$ factors describe the back action of the non-condensate on the evolution of mode `p'.

Although it is physically obvious that condensate number fluctuations can change the energy of an excitation it is perhaps less clear that phase fluctuations can have a similar effect. In this regard we note that a global change in the phase of both the condensate and all the excited states has of course no physical consequences. If we put $\lambda_2(t) \rightarrow \lambda_2(t) + d\lambda$ where $d\lambda$ is a constant, then it is easy to see from Eqs.~(\ref{TDGGPE}), (\ref{def_uvrt}) and (\ref{uvt}) that $\Phi_2\rt$, $u_i\rt$ and $v_i\rt$ transform according to $\Phi_2\rt \rightarrow \Phi_2\rt e^{id\lambda t/\hbar}$, $u_i\rt \rightarrow u_i\rt e^{id\lambda t/\hbar}$ and $v_i\rt \rightarrow v_i\rt e^{-id\lambda t/\hbar}$. Comparison with Eqs.~(\ref{TDGGPE}), (\ref{nt_qp}) and (\ref{mt_qp}) shows that the additional phase factor cancels in the GGPE. However, the perturbation does not necessarily cause the same phase fluctuations in the condensate and the excited states so there may be \emph{relative} phase fluctuations which are physically meaningful and which can affect the energy. For example, the term $P_{00}(\omega)$ in the Beliaev and Landau perturbation matrix elements of Eqs.~(\ref{PBij}) and (\ref{PLij}) comes from condensate phase fluctuations [c.f. Eq.~(\ref{dl0})]. This term is subtracted from the perturbation driving the excited states so that only the relative phase of the condensate and non-condensate has any physical effect.

As shown by Eq.~(\ref{c_coeffs}) and noted below Eq.~(\ref{dl0}), the $c_i$ coefficients are only non-zero for modes with the same symmetry as the condensate. This means that condensate phase and number fluctuations will only have an effect on the response if one or both of the modes `p' and `q' has the appropriate symmetry. Similarly, the effects of $\delta f\rw$ and the orthogonal projector $\hat{Q}$ in the $Y_{pij}/Y_{qij}$ coefficients are only relevant for subspaces where either $i$ or $j$ has the symmetries of the condensate. These spaces form a small subset of the whole, which means that these contributions describe finite size effects and should be negligible in the thermodynamic limit. Nonetheless, they can be significant for the finite systems of recent experiments. This is discussed further in \cite{Morgan03c,Morgan03d} where we consider the numerical implementation of the current theory. All these effects vanish in the homogeneous limit where the $c_i$ coefficients are identically zero (see Sec.~\ref{sec_homogeneous}). As far as we know, the contribution to the dynamic shift from the $c_i$'s has not been obtained previously.

The quantity $\Delta P_{p0}^{(D)}(\omega)$ given in Eq.~(\ref{dPp}) describes the excitation of the non-condensate by the perturbation and the subsequent coupling to the condensate. The structure of $\Delta P_{p0}^{(D)}(\omega)$ is similar to that of $\Delta E_{pq}^{(D)}(\omega)$ and the physical processes involved are the same, the only difference being how the non-condensate fluctuations are created. In this case, the matrix element $P_{ij}^{(B)}(\omega)$ describes the creation of pairs of quasiparticles by the perturbation while $P_{ij}^{(L)}(\omega)$ describes the rearrangement of existing quasiparticles. The $Y_{pij}$ factors describe the subsequent coupling to the condensate. However, $\Delta P_{p0}^{(D)}(\omega)$ has no contribution from the condensate modes and has no UV renormalization.

We note that the factor of $1/\sqrt{N_0}$ in the definition of $\Delta P_{p0}^{(D)}(\omega)$ in fact cancels with a factor of $\sqrt{N_0}$ contained in the definition of the $Y_{pij}$ coefficients [see Eqs.~(\ref{Ys})-(\ref{Jij}) and Eq.~(\ref{c_coeffs})]. This means that $\Delta P_{p0}^{(D)}(\omega)$ has no explicit dependence on $N_0$, although of course it still has an implicit dependence on this quantity via the quasiparticle energies and wavefunctions from which it is constructed. It is therefore expected to be only weakly-dependent on $N_0$ for a given temperature.

\section{Discussion} \label{sec_discussion}

\subsection{Homogeneous limit} \label{sec_homogeneous}

The theory presented above simplifies considerably in the homogeneous, thermodynamic limit where a number of its properties can be studied analytically. In this limit, the static solutions to the ordinary and generalized GPE are simply $\cond = \gcond = 1/\sqrt{V}$ where $V$ is the volume of the system, so $\Delta \Phi_{20}(\bfr) = 0$ and there are no condensate shape effects. The static quasiparticle solutions of Eq.~(\ref{LBdG_matrix}) are simply plane waves $\mathcal{X}_{\mathbf{k}}(\bfr) \propto e^{i\mathbf{k.r}}$ with the famous Bogoliubov dispersion relation $\epsilon_k = [(c\hbar k)^2+(\hbar^2k^2/2m)^2]^{1/2}$, where the speed of sound is $c^2 = n_0U_0/m$. The parameters $c_i$ defined in Eq.~(\ref{c_coeffs}) are zero for $i \neq \pm 0$ and therefore the $Y_{pij}^{(A)}$, $Y_{pij}^{(B1)}$ and $Y_{pij}^{(B2)}$ coefficients defined in Eq.~(\ref{Ys}) simplify to the integrals $A_{pij}$, $B1_{pij}$ and $B2_{pij}$ of Eqs.~(\ref{A})-(\ref{B2}). It also follows that the equilibrium value of $f(\bfr)$ is zero. Using the expression derived by Giorgini and co-authors for the variance in the condensate population \cite{Giorgini98a}, it is straightforward to show that terms involving $\Delta N_0$, vanish in the thermodynamic limit. The energy shift $\Delta E_f^{(S)}(p,q)$ of Eq.~(\ref{defs}) and the contribution from the condensate modes to the dynamic shift, $\Delta E_{0}(p,q)$ of Eq.~(\ref{de0}) also vanish in this limit. Thus the additional terms introduced by the number-conserving approach are finite size effects and are not relevant in the homogeneous, thermodynamic limit. The theory therefore reduces in this case to the Beliaev-Popov theory which has been discussed by a number of authors \cite{Beliaev58,Mohling60,Shi98,Fedichev98,Giorgini00,Morgan00}.

The remaining expressions for the static and dynamic energy shifts can be calculated explicitly. The excited states form a continuum, so the self-energy $\Sigma_p(\omega+i\gamma)$ is only weakly dependent on frequency over the width of a resonance and excitation energies and decay rates can be calculated using Eqs.~(\ref{E_p}) and (\ref{Gamma_p}) and taking the limit $\gamma \rightarrow 0^+$. The Beliaev and Landau decay rates calculated in this way generate all the standard results in the literature \cite{Beliaev58,Mohling60,Pitaevskii97,Fedichev98,Shi98,Giorgini98,Giorgini00,Morgan00} and are proportional to the small parameter of Eq.~(\ref{validity}). 

When the energy shifts are calculated it is found that the static and dynamic terms are both large, each giving an infra-red divergent contribution to the energy of the form $\Delta E_k^{(S)} \sim -\Delta E_k^{(D)} \propto 1/k$ as $k \rightarrow 0$. However, as shown explicitly in \cite{Fedichev98,Giorgini00,Morgan00}, the divergence in the static part is exactly cancelled by the divergence in the dynamic part and the resulting energy shift is finite, gapless and proportional to the parameter of Eq.~(\ref{validity}). The validity condition for the theory then comes from requiring that the energy shifts and decay rates are small compared with the unperturbed Bogoliubov energies.

\subsection{Non-condensate collisions} \label{sec_non-condensate_collisions}

The theory developed in Sec.~\ref{LINEAR_GGPE} represents a systematic treatment of a dilute Bose gas in the collisionless limit when the validity conditions of Eq.~(\ref{validity}) are satisfied. As the density and temperature increase, however, we will eventually reach a regime where higher order terms become important and ultimately the theory will fail in the hydrodynamic limit. The most important physical processes we have neglected are collisions of two non-condensate atoms where each remains in the thermal cloud. Of course a rigorous treatment of this process would also have to include a variety of other effects of the same order \cite{Walser99}. In this section we simply estimate the importance of such collisions in the homogeneous limit and show that they are of higher order than we consider in the small parameter of the theory.

We can estimate the non-condensate collision rate at temperatures greater than the chemical potential using classical kinetic theory. In this case, the collision rate $\Gamma_{\text{nc}}$ is given by
\beq
\Gamma_{\text{nc}} \sim \nt\sigma\bar{v},
\eeq
where $\sigma = 8\pi a_s^2$ is the scattering cross-section and $\bar{v} = (8k_{\text{B}}T/m\pi)^{1/2}$ is the mean speed. The non-condensate density can be approximated by the critical density $\nt \approx n_{\text{crit}} = \zeta(3/2)/\lambda_{\text{dB}}^3$ where $\lambda_{\text{dB}} = (2\pi\hbar^2/mk_{\text{B}}T)^{1/2}$ is the thermal de Broglie wavelength and $\zeta(x)$ is the Riemann zeta function with $\zeta(3/2) \approx 2.612$. Combining these results gives
\beq
\Gamma_{\text{nc}} \sim \frac{\zeta(3/2)8ma_s^2(k_{\text{B}}T)^2}{\pi\hbar^3}.
\eeq

The principal assumption of our treatment of non-condensate dynamics is that quasiparticles form a good basis, i.e. they evolve mainly under the influence of the condensate mean-field. This means that the time-scale for quasiparticle collisions must be long compared to their frequency at the relevant energy scale. As the dominant contribution to the self-energy comes from quasiparticles near the chemical potential $\mu=n_0U_0$, the importance of non-condensate collisions is described by the dimensionless parameter
\beq
\frac{\hbar \Gamma_{\text{nc}}}{n_0U_0} \sim 32\zeta(3/2)\left (\frac{k_{\text{B}}T}{n_0U_0} \right) ^2 (n_0a_s^3).
\eeq
Comparison with Eq.~(\ref{validity}) shows that this is proportional to the square of the small parameter of the theory, proving that non-condensate collisions are only relevant at the order beyond the current calculation.

\subsection{The dipole mode} \label{sec_dipole_mode}

The dipole mode (also known as the Kohn mode) corresponds to a centre-of-mass oscillation of the atomic cloud, with the condensate and non-condensate oscillating in phase. In a harmonic trap, this means that the atoms see a time-dependent potential which is a linear function of position across the cloud, corresponding to a spatially uniform force. Every atom therefore undergoes the same acceleration so that the relative motion of a pair of atoms is entirely unaffected. The evolution in the centre-of-mass frame is consequently the same as if the cloud was stationary, completely independent of temperature or particle interactions (this is the essential physical content of the generalized Kohn theorem) \cite{Dobson94}. For this reason the dipole modes are a very useful experimental diagnostic for determining the trap frequencies, but they are also an important test of the theoretical description of non-condensate dynamics. In this section we show that our linear response treatment describes the dipole modes correctly to within the order of the calculation. For this to be the case, it is important to include the effect of the external perturbation on the non-condensate, via the static and dynamic shifts $\Delta P_{p0}^{(S)}(\omega)$ and $\Delta P_{p0}^{(D)}(\omega)$. We focus on oscillations along the $z$-axis but similar comments apply to the dipole modes in the $x$-$y$ plane.

\subsubsection{The dipole mode in the GPE}

As is well-known, the dipole mode is obtained exactly in the GPE of Eq.~(\ref{TDGPE}) \cite{Fetter98}. It is easy to show that for a harmonic potential, the time-dependent GPE has exact solutions consisting of eigenstates oscillating along a principal axis of the trap with no change of shape \cite{Morgan97}. A consequence of this is that the BdG equations for the excitations, Eq.~(\ref{LBdG_matrix}), have solutions with energies corresponding to the principal trap frequencies.

We can generate dipole oscillations by varying the position of the centre of the harmonic trap. For oscillations along the $z$-axis we define $P\rt$ by
\beq
V\rt = V(\bfr) + P\rt = V(x,y,z-z_p),
\label{dipole_Vrt}
\eeq
where $z_p(t)$, the position of the trap minimum in the $z$-direction, is an arbitrary function of time. For a harmonic trap of frequency $\omega_z$ in the $z$-direction, the spatial dependence of the perturbation is therefore given by
\beq
P\rt = -m\omega_z^2zz_p(t).
\label{dipole_Prt}
\eeq

The condensate responds to this perturbation by oscillating along the $z$-axis with no change in shape. We can therefore write the time-dependent condensate wavefunction as
\beq
\Phi_0 \rt = \Phi_0(x,y,z-z_0)e^{ikz},
\eeq
where $\Phi_0 (\bfr)$ is an eigenstate solution of the time-independent GPE while $z_0(t)$ and $\hbar k(t)$ describe the centre-of-mass displacement and momentum respectively. We note that in general the condensate position does not coincide with the centre of the trap, i.e. $z_0(t) \neq z_p(t)$. The extra energy associated with the oscillation changes the rate of condensate phase rotation and can be dealt with by choosing the parameter $\lambda_0(t)$ via $\lambda_0(t) \rightarrow \lambda_0 + \delta \lambda_0(t)$. For a harmonic potential, these expressions provide a solution to the full time-dependent GPE provided that $z_0(t)$, $k(t)$ and $\delta \lambda_0(t)$ satisfy
\beq
\begin{aligned}
\frac{d^2 z_0}{dt^2} &= -\omega_z^2(z_0 - z_p), \quad \frac{\hbar k}{m} = \frac{dz_0}{dt},\\
\delta \lambda_0 &= \frac{\hbar^2 k^2}{2m} + \frac{1}{2}m\omega_z^2(z_p^2-z_0^2).
\end{aligned}
\label{dipole_z0eqns}
\eeq
The dipole mode is therefore obtained exactly in the GPE theory, regardless of the amplitude of the oscillations.

If we now consider the linear limit we see that $\delta \lambda_0 \approx 0$ because it depends quadratically on small quantities. This is consistent with the general expression for $\delta \lambda_0$ given in Eq.~(\ref{dl0}), which is only non-zero for modes with the same symmetry as the condensate. We then have
\beq
\delta \Phi_0 \rt = -z_0(t) \frac{\partial \cond}{\partial z} + ik(t)z\cond.
\label{dPhi0_dipole}
\eeq
Using the GPE it is straightforward to obtain equations for $\partial \cond/\partial z$ and $z\cond$ and from these to show that Eq.~(\ref{LBdG_matrix}) has the exact solutions \cite{Fetter98}
\beq
\mathcal{X}_{D\pm}(\bfr) = \frac{N_D}{2} \begin{pmatrix} \displaystyle \frac{\partial \cond}{\partial z} \mp z\cond/z_{\text{osc}}^2 \\[2mm] \displaystyle \frac{\partial \Phi_0^*(\bfr)}{\partial z} \pm z\Phi_0^*(\bfr)/z_{\text{osc}}^2\end{pmatrix},
\label{dipole_BdGuv}
\eeq
where $z_{\text{osc}} = (\hbar/m\omega_z)^{1/2}$ is the oscillator length unit in the $z$-direction and $N_D$ is a normalization constant. These are the positive and negative energy dipole modes with frequencies $\omega_{D\pm} = \pm \omega_z$. We note that the orthogonal projectors play no role in the BdG equations for these modes because of parity considerations.

The condensate fluctuation spinor corresponding to Eq.~(\ref{dPhi0_dipole}) can therefore be written as
\begin{align}
\mathcal{D}\Phi_0\rw = &-\frac{1}{N_D} \left( z_0 + ikz_{\text{osc}}^2 \right) \mathcal{X}_{D+}(\bfr) \nonumber \\
&- \frac{1}{N_D} \left( z_0 - ikz_{\text{osc}}^2 \right) \mathcal{X}_{D-}(\bfr).
\label{dipole_expansion}
\end{align}
Solving the equations for $z_0$ and $k$ in the Fourier domain gives
\beq
z_0(\omega) \pm ik(\omega)z_{\text{osc}}^2 = \frac{\mp\omega_z z_p(\omega)}{\omega\mp\omega_z}.
\eeq
A general oscillation of the condensate centre-of-mass therefore consists of a linear superposition of the positive and negative dipole modes with coefficients which are Lorentzians centred on $\pm \omega_z$ and weighted by the frequency spectrum of the perturbation.

\subsubsection{The dipole mode in the generalized GPE}

The nonlinear term in the GPE plays no role in the dipole mode because interactions are unaffected by centre-of-mass translations. This is true regardless of the particular functional form of the nonlinearity, provided it depends only on internal coordinates of the condensate. Denoting a general nonlinear term by $W\rt$, all we require for the dipole mode is that this function translates like the condensate, i.e that it evolves according to
\beq
W\rt = W(x,y,z-z_0)e^{ikz},
\label{translate_W}
\eeq
where $W (\bfr)$ indicates the static form. The fact that the shape of the oscillating condensate satisfies the time-independent GGPE in terms of $W (\bfr)$ is then sufficient to remove this term from the equation of motion.

The existence of an exact dipole mode in the GGPE is therefore assured if we can write this equation in terms of some function $W\rt$ which obeys Eq.~(\ref{translate_W}). We therefore define $W\rt$ by the last three lines of Eq.~(\ref{TDGGPE}) with UV renormalization, i.e. 
\begin{align}
W\rt = (N_0 + \Delta N_0)U_0|\Phi_2\rt|^2 \Phi_2\rt& \\
+ 2U_0\nt\rt\Phi_2\rt + U_0\mt^{\text{R}}\rt\Phi_2^*\rt&-f\rt. \nonumber
\end{align}
This obeys Eq.~(\ref{translate_W}) if $\nt\rt$, $\mt^{\text{R}}\rt$ and $f\rt$ evolve according to
\beq
\left.
\begin{aligned}
\nt\rt &= \nt(x,y,z-z_0),\\
\mt^{\text{R}}\rt &= \mt^{\text{R}}(x,y,z-z_0)e^{2ikz},\\
f\rt &= f(x,y,z-z_0)e^{ikz},
\end{aligned}
\right \}
\eeq
and these in turn are assured if the Bogoliubov spinors evolve as
\beq
\mathcal{X}_i\rt = \bpm e^{ikz} & 0 \\ 0 & e^{-ikz} \epm \mathcal{X}_i(x,y,z-z_0)e^{-i\epsilon_i t/\hbar}.
\label{dipole_Xrt}
\eeq

The static functions $\mathcal{X}_i(\bfr)$ are defined by Eq.~(\ref{LBdG_matrix}) and depend on $\cond$, the static solution to the ordinary GPE, while the time-dependent functions $\mathcal{X}_i\rt$ evolve according to Eq.~(\ref{uvt}). It is easy to show from Eq.~(\ref{uvt}), that if the perturbation has the form of Eq.~(\ref{dipole_Prt}), the condensate wavefunction is taken to be $\Phi_0 (x,y,z-z_0)e^{ikz}$ and $z_0$, $k$ and $\delta \lambda$ satisfy Eq.~(\ref{dipole_z0eqns}), then Eq.~(\ref{dipole_Xrt}) is satisfied. In this case $W\rt$ satisfies Eq.~(\ref{translate_W}) and we have an exact solution to the time-dependent GGPE
\beq
\Phi_2 \rt = \Phi_2(x,y,z-z_0)e^{ikz},
\label{dipole_ggpe}
\eeq
where $\Phi_2(\bfr)$ is an eigenstate solution. We note that it is important to include the effect of the perturbation $P\rt$ on the non-condensate for this argument to work correctly. 

This is not quite the end of the matter, however, for two reasons. First, we solve the static GGPE by linearizing the change in shape and energy around $\cond$ and $\lambda_0$ as in Eqs.~(\ref{dPhi20}) and (\ref{dlambda20}). This means that the above argument only applies to order $\Delta \Phi_{20}(\bfr)$. Second, the proof of Eq.~(\ref{dipole_Xrt}) requires that $\mathcal{X}_i\rt$ is driven by an oscillating condensate with the same shape as was used to define the static values, i.e. $\cond$. This requires $\cL_2\rt$ of Eq.~(\ref{bigL2rt}) to be written in terms of $\Phi_0\rt$ rather than $\Phi_2\rt$. Nonetheless, the use of $\Phi_2\rt$ here is essential for calculating general energy shifts correctly, as discussed in Appendix~\ref{app_phi2_in_L}. For centre-of-mass oscillations, however, the difference between the two wavefunctions is proportional to the change in shape of the static condensate. This is a small correction of order the small parameter of the theory and the resulting error is therefore of higher order than we consider.

The above discussion indicates that the dipole modes are not obtained exactly in the current theory, but the error will be beyond the order of the calculation. This can be confirmed by considering the linearized equations we actually solve, i.e. Eq.~(\ref{dPhi2w_matrix}). If Eq.~(\ref{dipole_Xrt}) is valid, this equation can be simplified considerably using the linearized expression for $\Delta \Phi_{20}(\bfr)$ given in Eq.~(\ref{app_dPhi20}). The result is that the error in the dipole mode is of order $(\Delta \Phi_{20})^2$, which is indeed beyond the order of the calculation. Interestingly, we also find that if we neglect the difference between $\Phi_0\rt$ and $\Phi_2\rt$ in high order terms, but retain the distinction in the leading order terms [i.e. the first line of Eq.~(\ref{dPhi2w_matrix})] then the dipole modes are obtained exactly.

This analysis applies to the case that we use a full basis expansion to evaluate $\delta \Phi_2\rt$, which requires solving the matrix problem of Eq.~(\ref{bp_GGPE_matrix}). In practice, the calculation is simplified significantly by assuming that only a single quasiparticle mode is excited. This corresponds to writing $\mathcal{D}\Phi_2\rw = b_{D+}(\omega)\mathcal{X}_{D+}(\bfr)$, with $\mathcal{X}_{D+}(\bfr)$ as in Eq.~(\ref{dipole_BdGuv}). In this case we find that the error in the dipole mode is now of order $\Delta \Phi_{20}$ but is localized around $\omega = -\omega_z$ and so should be very small in the frequency range of interest, namely $\omega \sim \omega_z$. Numerical results confirming the accurate calculation of the dipole modes in our formalism will be given elsewhere \cite{Morgan03c}.

In summary, a full implementation of the theory described in this paper will obtain the dipole modes correctly to the order of the calculation, with errors of order $(\Delta \Phi_{20})^2$. If we additionally assume that only a single quasiparticle mode is needed to describe the oscillations then the error is of order $\Delta \Phi_{20}$ but with a small prefactor at the frequencies of interest. In both cases, it is important to include the effect of the external perturbation on the non-condensate.

\section{Conclusions}

In conclusion, we have presented a theory of the linear response of Bose-Einstein condensates to external perturbations at finite temperature. We use a full quasiparticle description of the non-condensate and its dynamic coupling to the condensate. An important physical process, usually neglected in analytical calculations, is the direct excitation of the non-condensate by the perturbation and its subsequent coupling to the condensate. This process is explicitly included in our analysis and was shown in \cite{Morgan03a} to be responsible for the anomalous behaviour of the $m=0$ excitation observed in the 1997 JILA experiment \cite{Jin97}. It is also necessary to include this process for a correct description of the dipole modes. 

The theory is based on the time-dependent, number-conserving equations of Castin and Dum \cite{Castin98}. This formalism describes a number of finite size corrections compared with the more usual symmetry-breaking approach, and these can be significant in calculations for recent experiments \cite{Morgan03a}. The dynamic coupling between condensed and non-condensed atoms is treated using perturbation theory in a quasiparticle basis. Despite statements to the contrary in the literature \cite{Reidl00,Jackson02}, a perturbative calculation is appropriate provided that care is taken to account for finite size effects and direct excitation of the non-condensate.

The theory developed here is a systematic, gapless extension of
the Bogoliubov theory which is valid in the collisionless limit of well-defined quasiparticles. It is consistent with the generalized Kohn theorem for the dipole modes and includes the time-dependent normal and anomalous averages, Beliaev and Landau processes, and all relevant finite size effects. Details on the numerical implementation of the theory for trapped gases will be given elsewhere \cite{Morgan03c}.

\begin{acknowledgments}
It is a pleasure to thank Keith Burnett, Matthew Davis, Simon Gardiner, David Hutchinson, Mark Lee, and Martin Rusch for numerous invaluable discussions. I am grateful to the Royal Society of London for financial support.
\end{acknowledgments}

\appendix

\section{Energy shifts} \label{app_energy_shifts}

In this appendix we give some details of the derivation of the formulae for the static and dynamic energy
shifts $\Delta E_{pq}^{(S)}$ and $\Delta E_{pq}^{(D)}(\omega)$ and the changes
in the excitation matrix elements $\Delta P_{p0}^{(S)}(\omega)$ and $\Delta
P_{p0}^{(D)}(\omega)$ quoted in the main text. We also give expressions for the change in energy and shape of the equilibrium condensate wavefunction. We use the words `static' and `dynamic' for energy shifts to refer to the role
of the non-condensate. In the linearization procedure, static shifts therefore arise from terms involving
the condensate fluctuation $\delta \Phi_{2} \rw$ and equilibrium
non-condensate mean fields, while dynamic shifts involve the static
condensate wavefunction and fluctuating non-condensate mean fields.

In this appendix we will frequently drop space, time and frequency dependence in intermediate stages of the calculation. Time/frequency-dependent quantities can be interpreted directly in the time-domain; in the frequency domain, complex-conjugated terms should be evaluated at $-\omega$, while unstarred quantities are evaluated at $+\omega$.

\subsection{Change in condensate energy and shape} \label{app_condensate_statics}

The change in the condensate energy and shape between the GPE and GGPE descriptions are given by $\Delta\lambda_{20}$ and $\Delta\Phi_{20}(\bfr)$ respectively. For consistency with the linear response treatment of the fluctuations, we calculate these quantities by linearizing around the static solution to the ordinary GPE. Substituting Eqs.~(\ref{dPhi20}) and (\ref{dlambda20}) into Eq.~(\ref{TIGGPE}) leads to
\beq
\cLGP_0 \bpm \Delta\Phi_{20}\\ \Delta\Phi_{20}^*\epm = \Delta\lambda_{20}\mathcal{X}_{\theta_0} - \bpm V_{\text{GGPE}} \\ -V_{\text{GGPE}}^{*} \epm,
\label{app_dPhi20}
\eeq
where $V_{\text{GGPE}}$ denotes all terms in the static GGPE which are not also in the GPE [approximated with $\gcond \rightarrow \cond$], i.e.
\begin{align}
V_{\text{GGPE}}(\bfr) = 2U_0\nt(\bfr)\cond + U_0\mt^{\text{R}}(\bfr)\ccond& \nonumber\\
-f(\bfr) + \Delta N_0U_0|\cond|^2\cond&.
\end{align}
Equation~(\ref{app_dPhi20}) is solved using an expansion in the quasiparticle basis of Eq.~(\ref{LBdG_matrix})
\beq
\bpm \Delta\Phi_{20}\\ \Delta\Phi_{20}^*\epm = \sum_{i\neq\pm0} b_i^{(s)}\mathcal{X}_i(\bfr),
\eeq
where the static coefficients $b_i^{(s)}$ are given by
\beq
b_i^{(s)} = -\frac{\chi_i}{\epsilon_i}\intdr \bigl [ u_i^*V_{\text{GGPE}}+v_i^*V_{\text{GGPE}}^* \bigr ].
\eeq
The change in the condensate eigenvalue is given by
\begin{multline}
\Delta\lambda_{20} = \intdr \Bigl [ \Bigr. \Phi_0^* V_{\text{GGPE}} \\
 + N_0U_0|\Phi_0|^2 \left( \Phi_0^* \Delta\Phi_{20} + \Phi_0\Delta\Phi_{20}^* \right) \Bigl. \Bigr ].
\end{multline}

A complication which arises in the linearization leading to Eq.~(\ref{app_dPhi20}) is the fact that $\Delta\lambda_{20}$ is not small (i.e. it is not proportional to the small parameter of the theory) except at zero temperature. The reason is that the non-condensate density $\nt(\bfr)$ has a large contribution which is simply the ideal-gas density. In fact we can write $\nt(\bfr)$ in the form $\nt(\bfr) = \nt_{\text{id}}(\bfr)+ \nt_{\text{int}}(\bfr)$, where $\nt_{\text{id}}(\bfr)$ is the ideal-gas contribution and $\nt_{\text{int}}(\bfr)$ is the part due to interactions, which is proportional to the small parameter of the theory. Although $\nt_{\text{id}}(\bfr)$ is large at finite temperature, it is also roughly constant over the spatial extent of the condensate (this is exactly true in the Thomas-Fermi and homogeneous limits). As a result, the change in shape of the condensate $\Delta \Phi_{20}(\bfr)$ is still proportional to the small parameter of the theory and the linearization procedure is valid.

The large size of $\Delta\lambda_{20}$ does not cause problems elsewhere in the theory because it is cancelled straightforwardly by other terms. We can see this explicitly if we assume that $\nt_{\text{id}}(\bfr)$ is in fact a constant. Its contribution to $\Delta\lambda_{20}$ is then simply $2U_0\nt_{\text{id}}$ and comparison with Eqs.~(\ref{de4}) and (\ref{delambda}) shows that the effect of $\nt_{\text{id}}(\bfr)$ in $\Delta E_{4}(p,q)$ cancels exactly with the corresponding contribution in $\Delta E_{\lambda}(p,q)$. Thus only $\nt_{\text{int}}(\bfr)$ and finite size effects in $\nt_{\text{id}}(\bfr)$ are relevant for excitations and a linear approximation remains appropriate.

\subsection{Static shifts} \label{app_static_shifts}

With the exception of the last two lines of Eq.~(\ref{deshape}) for $\Delta E_{\text{shape}}(p,q)$
and the expression for $\Delta E_{f}^{(S)}(p,q)$ of Eq.~(\ref{defs}), the static shifts are derived
straightforwardly by calculating the overlap integral of the third line of
Eq.~(\ref{dPhi2w_matrix}) with the quasiparticle wavefunctions for mode `p' using Eq.~(\ref{project_b}) and multiplying by $\chi_p^2 = 1$. This
gives $\Delta E_{4}(p,q)$, $\Delta E_{\lambda}(p,q)$, $\Delta E_{\Delta N_0}(p,q)$ and the first two lines
of $\Delta E_{\text{shape}}(p,q)$ directly. The last two lines of Eq.~(\ref{deshape}) come partly from condensate phase fluctuations and partly from the condensate norm mode, as we now show.

Explicitly, the second term of Eq.~(\ref{dPhi2w_matrix}) is
\begin{align}
\cLGP_0 \mathcal{D}\Phi_2\rw &= 2N_0U_0|\Phi_0|^2 \mathcal{X}_{\theta_0}b_{N_0}(\omega) \nonumber \\
&+\sum_{i\neq\pm0} \left[ \epsilon_i\mathcal{X}_i + c_i\mathcal{X}_{\theta_0} \right]b_i(\omega),
\end{align}
where $b_{N_0}(\omega)$ is given in Eq.~(\ref{b0R}). The contribution to mode `p' from the first line of Eq.~(\ref{dPhi2w_matrix}) is therefore
\beq
(\hbar \omega - \epsilon_p) b_p(\omega) - 2\chi_p c_p^* b_{N_0}(\omega) - \chi_p P_{p0}(\omega),\label{app_bp1}
\eeq
where $c_p$ and $P_{p0}(\omega)$ are given by Eqs.~(\ref{c_coeffs}) and (\ref{P_p0}) respectively.

The contribution from the second line of Eq.~(\ref{dPhi2w_matrix}) is
\beq
\chi_p \intdr \bigl [ P\rw -\delta\lambda_2^r (\omega) \bigr ] \bigl  [ u_p^* \Delta\Phi_{20} + v_p^* \Delta\Phi_{20}^* \bigr ].
\eeq
In this expression, we can use the leading order result for $\dl_2^r(\omega) = \dl_0(\omega)$ given by Eq.~(\ref{dl0}), which removes condensate phase fluctuations. Combining the result with Eq.~(\ref{app_bp1}), multiplying by $\chi_p^2 = 1$ and regrouping we get
\begin{align}
\chi_p ( \hbar \omega &- \epsilon_p ) b_p(\omega) - P_{p0}(\omega) = \nonumber\\
&\Delta P_{p0}^{(S)}(\omega) + \sum_{q\neq\pm0} \Delta E_{\perp}(p,q) b_q(\omega) + \ldots,
\end{align}
with $\Delta P_{p0}^{(S)}(\omega)$ as in Eq.~(\ref{DP_p0_s}) and $\Delta E_{\perp}(p,q)$ defined by
\begin{align}
\Delta E_{\perp}(p,q) = &-c_p^*\intdr \left[ \Delta\Phi_{20}^{*} u_q + \Delta\Phi_{20} v_q \right] \nonumber \\
& -c_q\intdr \left[ \Delta\Phi_{20} u_p^* + \Delta\Phi_{20}^{*} v_p^* \right].
\label{app_deortho}
\end{align}

$\Delta E_{\perp}(p,q)$ depends on the change in shape of the static condensate wavefunction and is therefore logically included in the definition of $\Delta E_{\text{shape}}(p,q)$, giving the last two lines of Eq.~(\ref{deshape}). As mentioned in the main text, in a time-independent, perturbative calculation this contribution arises from the change in the definition of the subspace orthogonal to the condensate when the condensate shape changes \cite{Morgan00}.

The final contribution to the static shifts comes from $\delta f\rw$ in the last line of Eq.~(\ref{dPhi2w_matrix}). The corresponding energy shift for mode `p' is given by
\beq
\Delta E_{f}(p) = -\intdr \left [u_{p}^{*} \delta f\rw + v_{p}^{*} \delta f^{*}\rmw \right ].
\label{app_def}
\eeq
Taking the expression for $f\rt$ given in Eq.~(\ref{def_f}) and linearizing we obtain
\beq
\delta f = \delta f^{(S)} + \delta f^{(D)},
\eeq
where the contribution from the static non-condensate is
\begin{align}
\delta f^{(S)} = U_0\intdrp \Bigl \{ \Bigr. & \bigl [ 2|\Phi_{0}|^2 \delta\Phi_2 + \Phi_0^{2}\delta \Phi_{2}^{*} \bigr ] \nt\rr \label{app_dfs}\\
 +&\bigl[ 2|\Phi_{0}|^2 \delta \Phi_2^{*} + \Phi_0^{*2} \delta\Phi_{2} \bigr] \mt \rr \Bigl. \Bigr \},\nonumber
\end{align}
 while the dynamic contribution has the form
\beq
\delta f^{(D)} = \intdrp U_0|\Phi_{0}|^2 \bigl [ \Phi_{0}\delta\nt\rr + \Phi_{0}^{*}\delta\mt\rr \bigr ].
\label{app_dfd}
\eeq 
All unspecified spatial coordinates are evaluated at $\bfrp$.

Using Eq.~(\ref{qp_expansion2}) with $b_{N_0} = b_{\theta_0} = 0$ correct to this order gives
\beq
\delta\Phi_2(\bfr) = \sum_{q \neq \pm 0} u_q(\bfr)b_q, \quad \delta\Phi_2^*(\bfr) = \sum_{q \neq \pm 0} v_q(\bfr)b_q,
\label{app_dPhi2_expansion}
\eeq
while $\nt\rr$ and $\mt\rr$ are obtained from the time-independent limit of Eqs.~(\ref{nt_qp}) and (\ref{mt_qp}). Substituting these results into Eqs.~(\ref{app_def}) and (\ref{app_dfs}) leads to an expression for the static part of the energy shift due to $f\rt$. We are left with integrals over $\bfrp$ of four wavefunctions which can be expressed in terms of the matrix elements $A_{qij}$, $B1_{qij}$ and $B2_{qij}$ defined in Eqs.~(\ref{A})-(\ref{B2}), and integrals over $\bfr$ of products of two functions which involve the factors $U_{ij}$, $V_{ij}$ and $W_{ij}$ of Eqs.~(\ref{Uij})-(\ref{Wij}). The final result can be written as
\beq
\Delta E_{f}^{(S)}(p) = -\sum_{q \neq \pm 0}  \Delta E_{f}^{(S)}(p,q)b_q (\omega),
\eeq
with $\Delta E_{f}^{(S)}(p,q)$ as in Eq.~(\ref{defs}).

\subsection{Dynamic shifts}\label{app_dynamic_shifts}

We now turn to the calculation of terms involving fluctuations of the
non-condensate. Following Blaizot and Ripka \cite{Blaizot_Ripka} we define a generalized density matrix $\mathcal{R}$ by
\begin{align}
\mathcal{R}(\bfr,\mathbf{r'},t) &= \left \langle \left [ \sigma_1 \bpm \Lambda (\mathbf{r'},t) \\
\Lambda^{\dagger} (\mathbf{r'},t) \epm \bpm \Lambda^{\dagger} \rt, & \Lambda \rt \epm \sigma_1
\right ]^{T} \right \rangle \nonumber \\
&= \bpm \nt(\bfr,\mathbf{r'},t) & \mt(\bfr,\mathbf{r'},t)  \\ \mt^*(\bfr,\mathbf{r'},t) & \quad \left [Q^* + \nt^* \right ](\bfr,\mathbf{r'},t) \epm.
\label{app_def_R}
\end{align}
Expanding the field operators as in Eq.~(\ref{Lqpexp}) and using the fact that the quasiparticle operators are time-independent and diagonal gives
\beq
\mathcal{R}(\bfr,\mathbf{r'},t) = \sum_i \mathcal{X}_i\rt \mathcal{X}_i^{\dagger}\rpt R_i,
\label{app_RX}
\eeq
where the sum is over all states and the coefficients $R_i$ are defined in terms of the time-independent quasiparticle populations $\{N_i\}$ by
\beq
R_i =
\begin{cases}
N_i, & (\chi_i = +1), \\
N_i+1, & (\chi_i = -1), \\
0, & (i = \pm 0).
\end{cases}
\label{app_Ri}
\eeq
We note that the time-dependent condensate modes give no direct contribution in this expression
because $R_{\pm 0} = 0$. It is nonetheless convenient to include them formally
in the summation as we have done in Eq.~(\ref{app_RX}). The reason is that when we expand the time-dependent wavefunctions in the time-independent basis [Eq.~(\ref{appdXexp})], the static condensate modes are needed to preserve orthogonality. Including them at the outset allows a symmetric treatment of the two summation indices which then appear.

We can write the time-dependent spinors in terms of small fluctuations around the static values given by the solutions of Eq.~(\ref{LBdG_matrix})
\beq
\mathcal{X}_i\rt = \bigl [ \mathcal{X}_i(\bfr) + \delta\mathcal{X}_i\rt \bigr ]e^{-i\epsilon_i t/\hbar}.
\eeq
Expanding the fluctuations in the static basis via
\beq
\delta\mathcal{X}_i\rt = \sum_j X_{ij}(t) \mathcal{X}_j(\bfr),
\label{appdXexp}
\eeq
gives, after linearizing
\beq
\delta \mathcal{R} = \sum_{ij} \mathcal{X}_j(\bfr) \mathcal{X}_i^{\dagger}(\mathbf{r'}) \left
[ R_i X_{ij} + R_j X_{ji}^* \right ].
\label{app_dRXpm}
\eeq

Using the time-dependent version of the orthonormality relation of Eq.~(\ref{orthosym}) and the
relationship between positive and negative norm spinors of Eq.~(\ref{X_minus_i}) we obtain the
following useful properties of the coefficients $X_{ij}$
\beq
\chi_j X_{ij} = -\chi_i X_{ji}^*, \qquad X_{ij} = X_{-i,-j}^*.
\label{app_Xijproperties}
\eeq
Using these results and the completeness relation of Eq.~(\ref{completeness}), we rewrite
Eq.~(\ref{app_dRXpm}) in terms of a sum over modes with positive norm only (including the state $+0$)
\begin{align}
\delta \mathcal{R} &=
\begin{aligned}[t]
\sum_{ij \geq 0} \frac{1}{2} \left [ \mathcal{X}_{-i}(\bfr)
\mathcal{X}_j^{\dagger}(\bfrp) + \mathcal{X}_{-j}(\bfr)
\mathcal{X}_{i}^{\dagger}(\bfrp) \right ]&\\
\times\left ( 1 + R_i + R_j \right )X_{i,-j}&
\end{aligned} \nonumber \\
&+\begin{aligned}[t]
\sum_{ij \geq 0} \frac{1}{2} \left [\mathcal{X}_i(\bfr) \mathcal{X}_{-j}^{\dagger}(\bfrp)
+ \mathcal{X}_j(\bfr) \mathcal{X}_{-i}^{\dagger}(\bfrp) \right ]&\\
\times\left ( 1 + R_i + R_j \right )X_{i,-j}^*&
\end{aligned} \nonumber \\
&+\begin{aligned}[t]
\sum_{ij \geq 0} \left [\mathcal{X}_i(\bfr)
\mathcal{X}_j^{\dagger}(\bfrp)
+ \mathcal{X}_{-j}(\bfr)\mathcal{X}_{-i}^{\dagger}(\bfrp) \right ]&\\
\times\left ( R_j -R_i \right )X_{ji} & 
\end{aligned} \nonumber \\
&+ \bpm 0 & 0 \\ 0 & \delta Q_2^*\rr \epm,
\end{align}
where
\beq
\delta Q_2^*\rr = - \bigl [ \Phi_0(\mathbf{r'}) \delta\Phi_2^*(\bfr) + \ccond \delta\Phi_2(\bfrp) 
\bigr ].
\label{app_dQ2}
\eeq

From this result, the formulae for $\delta\nt\rr$ and $\delta\mt\rr$ can be read off from the first row
\begin{align}
\delta\nt\rr &=
\begin{aligned}[t]
\sum_{ij \geq 0} \frac{1}{2}\left [ v_i^*(\bfr)u_j^*(\mathbf{r'}) + v_j^*(\bfr)u_i^*(\mathbf{r'}) \right ]&\\
\times\left ( 1 + R_i + R_j \right )X_{i,-j}&
\end{aligned} \nonumber \\
&+\begin{aligned}[t]
\sum_{ij \geq 0} \frac{1}{2}\left [ u_i(\bfr)v_j(\mathbf{r'}) + u_j(\bfr)v_i(\mathbf{r'}) \right ]&\\
\times\left ( 1 + R_i + R_j \right )X_{i,-j}^*&
\end{aligned} \nonumber \\
&+\begin{aligned}[t]
\sum_{ij \geq 0}  \left [u_i(\bfr)u_j^*(\mathbf{r'}) + v_j^*(\bfr)v_i(\mathbf{r'})\right ]&\\
\times\left ( R_j -R_i \right )X_{ji}&,
\end{aligned}
\label{app_dnrrw}
\end{align}
\begin{align}
\delta\mt\rr &=
\begin{aligned}[t]
\sum_{ij \geq 0} \frac{1}{2}\left [ v_i^*(\bfr)v_j^*(\mathbf{r'}) + v_j^*(\bfr)v_i^*(\mathbf{r'}) \right ]&\\
\times\left ( 1 + R_i + R_j \right )X_{i,-j}&
\end{aligned} \nonumber \\
&+\begin{aligned}[t]
\sum_{ij \geq 0} \frac{1}{2}\left [ u_i(\bfr)u_j(\mathbf{r'}) + u_j(\bfr)u_i(\mathbf{r'}) \right ]&\\
\times\left ( 1 + R_i + R_j \right )X_{i,-j}^*&
\end{aligned} \nonumber \\
&+\begin{aligned}[t]
\sum_{ij \geq 0}  \left [u_i(\bfr)v_j^*(\mathbf{r'}) + v_j^*(\bfr)u_i(\mathbf{r'})\right ]&\\
\times\left ( R_j -R_i \right )X_{ji}&. 
\end{aligned}
\end{align}

The dynamic terms give a contribution to the evolution of mode `p' which comes from the projection of the last two lines of Eq.~(\ref{dPhi2w_matrix})
\begin{align}
\Delta P_{p0}^{(D)}(\omega) + &\Delta E_p^{(D)}(\omega) = \nonumber\\
\intdr \biggl \{ \biggr. &2U_0 \delta\nt \bigl [ u_p^* \Phi_0 + v_p^* \Phi_0^* \bigr ] \label{app_DEd}\\
&+ U_0 \delta\mt^{\text{R}} u_p^* \Phi_0^* + U_0 \delta\mt^{\text{R}*} v_p^* \Phi_0 \nonumber\\
& - \frac{c_p^*}{N_0}\delta\nt - \Bigl [ u_p^* \delta f^{(D)} + v_p^* \delta f^{(D)*} \Bigr ] \biggl. \biggr \},\nonumber
\end{align}
where $\delta f^{(D)}$ is given in Eq.~(\ref{app_dfd}) and we have used the fact that $\delta N_0 = -\intdr \delta \nt$. Using the above equations for $\delta\nt\rr$ and $\delta\mt\rr$, this becomes
\begin{align}
\Delta P_{p0}^{(D)}(\omega) &+ \Delta E_p^{(D)}(\omega) = \nonumber \\
&\sum_{ij \geq 0} \frac{Y_{pij}^{(A)*}X_{i,-j}(\omega)}{2\sqrt{N_0}} \left [ 1 + N_{i} + N_{j} \right ] \nonumber\\
+ &\sum_{ij \geq 0} \frac{Y_{pij}^{(B1)*}X_{i,-j}^*(-\omega)}{2\sqrt{N_0}} \left [1 + N_{i} + N_{j} \right ]\nonumber\\
+ &\sum_{ij \geq 0} \frac{Y_{pij}^{(B2)*}X_{ji}(\omega)}{\sqrt{N_0}} \left [N_{j} - N_{i} \right ]\nonumber\\
+ & \Delta E^{\text{R}}(p,q),
\label{app_dynamicshift}
\end{align}
where the coefficients $Y_{pij}^{(A)}$, $Y_{pij}^{(B1)}$ and $Y_{pij}^{(B2)}$ are defined in the main text in Eq.~(\ref{Ys}). The UV renormalization $\Delta E^{\text{R}}(p,q)$ is given by Eq.~(\ref{deR}). This result is obtained straightforwardly from Eq.~(\ref{app_DEd}) using the contribution to $\delta \mt^{\text{R}}$ from the last part of Eq.~(\ref{rmt}).

It remains to find an expression for the coefficients $X_{ij}(\omega)$. This is achieved by solving the equation of motion for the time-dependent Bogoliubov functions $\mathcal{X}_i\rt$ given in Eq.~(\ref{uvt}). After linearization, this becomes
\beq
i \hbar \frac{\partial}{\partial t} \delta\mathcal{X}_i = \left[ \cL_0 -\epsilon_i
\right] \delta\mathcal{X}_i + \delta\cL_2 \mathcal{X}_i.
\label{app_uvt_linear}
\eeq
with $\cL_0$ as in Eq.~(\ref{bigL0r}). $\delta\cL_2$ is obtained by linearizing $\cL_2\rt$ of Eq.~(\ref{bigL2rt}) using
\beq
\left.
\begin{aligned}
\hsp\rt &= \hsp(\bfr) + P\rt,\\
\Phi_2 \rt &\approx \cond + \delta \Phi_2 \rt,\\
\lambda_2(t) &\approx \lambda_0 + \dl_{2}^r(t),\\
\hat{Q}_2(t) &\approx \hat{Q}_0 + \delta \hat{Q}_2(t),
\end{aligned} \right \}
\label{app_uvt_linearizations}
\eeq
with $\delta \hat{Q}_2(t)$ as in Eq.~(\ref{app_dQ2}). Note that, correct to the order of this calculation, we have used $\gcond \approx \cond$ and $\lambda_2 \approx \lambda_0$ for static condensate properties. We must retain the distinction between $\delta\Phi_0\rt$ and $\delta\Phi_2\rt$, however, for the reason given in Appendix~\ref{app_phi2_in_L}.

Substituting Eq.~(\ref{appdXexp}) into Eq.~(\ref{app_uvt_linear}), taking the Fourier transform and projecting out the required coefficient gives
\beq
X_{ij}(\omega) = \frac{\chi_j \intdr \left [ \mathcal{X}_j^{\dagger}(\bfr) \sigma_3 \delta\cL_2\rw
\mathcal{X}_i(\bfr) \right ]}{\hbar \omega + \epsilon_i - \epsilon_j}, \label{appXijw}
\eeq
applying for all states $i$ and $j$. It is easily verified that this expression satisfies the relations of Eq.~(\ref{app_Xijproperties}).

We obtain explicit expressions for $X_{ij}(\omega)$ by writing $\delta\cL_2\rw$ in terms of $P\rw$, $\dl_2^r(\omega)$, $\delta\Phi_2\rw$ and $\delta \hat{Q}_2(t)$. The effect of the orthogonal projector depends on whether or not it acts on the condensate modes in the expansion. The expression for the coefficients $X_{ij}(\omega)$ therefore depends on whether or not one of the labels $i$ or $j$ corresponds to $\pm 0$. For $i,j \neq \pm 0$ and $\chi_i,\chi_j = +1$, we can write $X_{ij}(\omega)$ in the form
\begin{subequations}
\begin{align}
\frac{X_{ji}(\omega)}{\sqrt{N_0}} &= \frac{ P_{ij}^L(\omega)/\sqrt{N_0} + \sum_{q\neq\pm0} Y_{qij}^{(B2)} b_q(\omega) }{\hbar \omega - \epsilon_i + \epsilon_j}, \\
\frac{X_{i,-j}(\omega)}{\sqrt{N_0}} &= \frac{ -\left [ P_{ij}^{B*}(-\omega)/\sqrt{N_0} + \sum_{q\neq\pm0} Y_{qij}^{(A)} b_q(\omega) \right ]}{\hbar \omega + \epsilon_i + \epsilon_j},  \\
\frac{X_{i,-j}^*(-\omega)}{\sqrt{N_0}} &= \frac{ P_{ij}^B(\omega)/\sqrt{N_0} + \sum_{q\neq\pm0} Y_{qij}^{(B1)} b_q(\omega) }{\hbar \omega - \epsilon_i - \epsilon_j}. 
\end{align}
\label{app_XYPs}
\end{subequations}
In deriving these expressions the following relations between the various coefficients we have defined are useful
\begin{subequations}
\begin{gather}
P_{ij}^B(\omega) = P_{-j,i}^{L*}(-\omega),\\
Y_{pij}^{(A)} = Y_{-p,ij}^{(B1)*} = Y_{p,-i,j}^{(B2)}.
\end{gather}
\end{subequations}
We also note that $Y_{pij}^{(B1)}$ is symmetric with respect to $i$ and $j$ while $Y_{pij}^{(A)}$ is invariant with respect to permutations of its indices. 

The form of $X_{ij}(\omega)$ for the case that one or both of the indices $i$ or $j$ refers to $\pm 0$ is different. The expression for $X_{ij}(\omega)$ in this case is most easily obtained using the fact that the time-dependent Bogoliubov functions $u_i\rt$ and $v_i\rt$ are orthogonal to the condensate, i.e. $\intdr \Phi_2^* \rt u_i\rt = \intdr \Phi_2 \rt v_i\rt = 0$. Using Eq.~(\ref{app_dPhi2_expansion}) for $i \neq \pm0$, $\chi_i = +1$ gives
\begin{align}
X_{i,+0} &= \intdr \Phi_0^* \delta u_i = - \intdr u_i \delta\Phi_2^* \nonumber\\
& = -\sum_{q\neq\pm0} W_{iq}b_q = -X_{+0,i}^*, \label{app_X0i}\\
X_{i,-0} &= \intdr \Phi_0 \delta v_i= -\intdr v_i \delta\Phi_2 \nonumber\\
& = -\sum_{q\neq\pm0} W_{qi}b_q = X_{+0,-i}, \label{app_X0mi}
\end{align}
which is consistent with Eq.~(\ref{app_Xijproperties}). The absence of condensate phase and norm modes (at leading order) gives
\beq
X_{+0,i = \pm 0} = X_{-0,i = \pm 0} = 0.
\label{app_X00}
\eeq
Equations~(\ref{app_X0i}) and (\ref{app_X0mi}) show that orthogonality between the time-dependent condensate and excited state wavefunctions requires $\delta u_i$ and $\delta v_i$ to have a non-zero overlap with the static condensate modes.

The expression for the dynamic shifts is finally obtained by substituting Eqs.~(\ref{app_XYPs})-(\ref{app_X00}) into Eq.~(\ref{app_dynamicshift}). The contribution to $X_{ij}(\omega)$ from the
perturbation $P \rw$ gives $\Delta P_{p0}^{(D)}(\omega)$, while $\Delta E_{p}^{(D)}(\omega)$ comes from the part of
$X_{ij}(\omega)$ involving the condensate fluctuations $\{b_q(\omega)\}$. In this latter contribution, we write the part involving the condensate mode separately, using Eqs.~(\ref{app_X0i}) and (\ref{app_X0mi}) which gives $\Delta E_{0}(p,q)$ in Eq.~(\ref{de0}). We note that the summations in $\Delta P_{p0}^{(D)}(\omega)$ of Eq.~(\ref{dPp}) exclude the mode $+0$ because, as in Eqs.~(\ref{app_X0i}) and (\ref{app_X0mi}), the terms this appears in can be written solely in terms of the condensate fluctuations.

The coefficients $X_{ij}(\omega)$ in Eq.~(\ref{app_XYPs}) are defined in terms of the same coefficients $Y_{qij}$ that appear in Eq.~(\ref{app_dynamicshift}) and are defined in Eq.~(\ref{Ys}). As discussed in the main text, the various contributions to these coefficients have a different physical origin in the two equations. The derivation of Eq.~(\ref{app_XYPs}) for $X_{ij}(\omega)$ shows that the term involving $c_q$ in the $Y_{qij}$'s comes from condensate phase fluctuations [the contribution from $\delta\lambda_2^r(\omega)$ to $\delta\cL_2\rw$]. The corresponding term in the $Y_{pij}$'s in Eq.~(\ref{app_dynamicshift}) comes instead from condensate number fluctuations [the term $-c_p^*\delta\nt/N_0$ in Eq.~(\ref{app_DEd})]. In addition, the terms involving $c_i$ and $c_j$ in the $Y$'s appear, in $X_{ij}(\omega)$, as a consequence of fluctuations in the orthogonal projector $\delta\hat{Q}_2$. The same contributions to the $Y$'s in Eq.~(\ref{app_dynamicshift}), come instead from $\delta f^{(D)}$. Finally, the matrix elements $A_{pij}$, $B1_{pij}$ and $B2_{pij}$ in the $Y$'s describe, in $X_{ij}(\omega)$, the driving of the non-condensate by the condensate fluctuation, while their appearance in Eq.~(\ref{app_dynamicshift}) describes the subsequent back action of the non-condensate on the condensate.

Finally, to calculate the condensate density fluctuations we need an expression for $\delta N_0(\omega) = -\intdr \delta \nt\rw$. Using Eqs.~(\ref{app_dnrrw}), (\ref{Iij}) and (\ref{Jij}) we obtain
\begin{align}
\delta N_0(\omega) = &-\sum_{ij \geq 0} \frac{J_{ij}^{*}X_{i,-j}(\omega)}{2} \left [ 1 + N_{i} + N_{j} \right ] \nonumber \\
&- \sum_{ij \geq 0} \frac{J_{ij}X_{i,-j}^*(-\omega)}{2} \left [1 + N_{i} + N_{j} \right ] \nonumber \\
&- \sum_{ij \geq 0} I_{ij}^{*}X_{ji}(\omega) \left [N_{j} - N_{i} \right ].
\label{app_dN0}
\end{align}
This expression can be broken into two parts, corresponding to the change in $\nt(\bfr)$ due to condensate fluctuations and the change due to the external perturbation. The resulting formulae are easily obtained using the above results for $X_{ij}(\omega)$.

\section{Effect of generalized GPE wavefunction in non-condensate equations} \label{app_phi2_in_L}

A subtle issue in the theoretical formalism concerns the appearance of the generalized GPE wavefunction $\Phi_2\rt$ rather than $\Phi_0\rt$ in the matrix $\mathcal{L}_2\rt$ which defines the evolution of the non-condensate [c.f. Eqs.~(\ref{bigL2rt})-(\ref{M2rt})]. In a self-consistent treatment of the non-condensate dynamics $\mathcal{L}_2\rt$ would also contain terms involving $\nt\rt$ and $\mt\rt$ describing the action of non-condensate fluctuations on themselves. Such a treatment has been given in \cite{Walser99} using a symmetry-breaking approach, and also includes additional kinetic terms in both the condensate and non-condensate equations of motion which are required for consistency. In this paper we treat the coupling between the condensate and non-condensate perturbatively so that the definition of $\mathcal{L}_2\rt$ given in Eqs.~(\ref{bigL2rt})-(\ref{M2rt}) is sufficient for our purposes.

However, the use of $\Phi_2\rt$ rather than $\Phi_0\rt$ in the equation of motion for the non-condensate produces some theoretical difficulties. For example, orthogonality between the condensate and excited states in the time-dependent case is only preserved to the order of the theory and is not exact. We stress that this is not a problem for the static basis we use to expand wavefunctions because this is defined by Eq.~(\ref{LBdG_matrix}) where the replacement $\gcond \rightarrow \cond$ is explicitly used, but the issue does arise when we solve for the dynamic fluctuations of the non-condensate, as in Appendix~\ref{app_dynamic_shifts}. Nonetheless, it is essential to use the generalized wavefunction in the dynamic case if we wish to obtain sensible expressions for energy shifts, as we demonstrate below. This should be valid within the linearized, perturbative approach of interest here, but a self-consistent treatment will be needed for higher order calculations or for situations where the condensate is perturbed beyond the linear regime.

For the purposes of this appendix, we distinguish between the condensate excitation coefficients $\{b_i(\omega)\}$ obtained from the GPE and GGPE theories by denoting the former by $\{b_i^{(0)}(\omega)\}$ and the later by $\{b_i^{(2)}(\omega)\}$. The evolution of the non-condensate introduces the dynamic energy shift $\Delta E_p^{(D)}(\omega)$ and the change in the excitation matrix element $\Delta P_{p0}^{(D)}(\omega)$ into the theory. Of these, only $\Delta E_p^{(D)}(\omega)$ depends on the treatment of the condensate dynamics. This quantity is written in Eq.~(\ref{DEd}) and depends on the expansion coefficients, which should be taken to be $\{b_i^{(2)}(\omega)\}$ rather than $\{b_i^{(0)}(\omega)\}$ because of the use of $\Phi_2\rt$ in the non-condensate evolution. Specializing to the case that only one quasiparticle mode is excited, this means that when we solve Eq.~(\ref{bp_GGPE}) $\Delta E_p^{(D)}(\omega)$ must be taken over to the left-hand-side of the equation (which contains $b_p^{(2)}$) and hence appears in the denominator of the resulting expression for $\mathcal{G}_p(\omega)$ given in Eq.~(\ref{G}). If instead we were to use $\Phi_0\rt$ to evolve the non-condensate then $\Delta E_p^{(D)}(\omega)$ would be defined in terms of the previously calculated coefficients $\{b_i^{(0)}(\omega)\}$ given in Eq.~(\ref{bp_GPE}). For consistency, we would also have to treat the static shifts of Eq.~(\ref{DEs}) in the same approximation (to avoid infra-red divergences). The result is that the self-energy $\Sigma_p(\omega)$ of Eq.~(\ref{sigmapw}) would appear in the numerator of $\mathcal{G}_p(\omega)$ rather than the denominator, and we would obtain
\begin{multline}
b_p^{(2)}(\omega) = \frac{\chi_p P_{p0}(\omega)}{\hbar \omega - \epsilon_p}\biggl [ \biggr. 1 + \frac{\Delta P_{p0}^{(S)}(\omega) + \Delta P_{p0}^{(D)}(\omega)}{P_{p0}(\omega)} \\
 + \frac{\Sigma_p(\omega)}{\hbar \omega - \epsilon_p}\biggl. \biggr ].
\label{app_bp0_GGPE_soln}
\end{multline}
Comparing this with Eqs.~(\ref{bp_GGPE_soln})-(\ref{G}), we see that it corresponds to writing
\beq
\mathcal{G}_p(\omega) = \frac{1}{(\hbar \omega - \epsilon_p)\left[1- \Sigma_p(\omega)/(\hbar \omega - \epsilon_p) \right]},
\eeq
and expanding the denominator to first order in $\Sigma_p(\omega)/(\hbar \omega - \epsilon_p)$. This is certainly not a valid approximation in the interesting region $\hbar \omega \sim \epsilon_p$, and does not lead to a shift in the peak of $\mathcal{G}_p(\omega)$ by an amount corresponding to $\Sigma_p(\omega)$. It is therefore necessary to include the condensate dynamics self-consistently in the theory to obtain sensible expressions for energy shifts and this is the reason why $\Phi_2\rt$ appears in the equation of motion for the non-condensate.

\section{Introduction of an imaginary frequency in the resolvent} \label{app_exp_resolution}

The expressions for the dynamic contributions to the energy shift $\Delta E_p^{(D)}(\omega)$ and excitation matrix element $\Delta P_{p0}^{(D)}(\omega)$ contain energy denominators which may vanish or become very small if a particular collision process is energetically allowed. To obtain finite quantities, essential for numerical work, it is therefore necessary to include a small imaginary part in the frequency via $\omega \rightarrow \omega + i \gamma$ \cite{Guilleumas99,Rusch00}. For a homogeneous gas, where an excitation couples to a continuum of decay channels, we can take the limit $\gamma \rightarrow 0^+$, but for trapped condensates the discrete nature of the states involved means that we must work with a finite (but small) value of $\gamma$. It is therefore necessary to justify its inclusion more carefully and to predict its approximate numerical value.

We can justify the inclusion of $\gamma$ by considering the finite experimental observation time. In fact, the coefficients $\{b_i(\omega)\}$ with $\gamma=0$ do not correspond to any experiments because they assume that the GPE describes the measured evolution of the condensate for all time. This of course is not the case and in any real experiment the condensate must first be created and allowed to settle into its equilibrium state. It is then excited by modulation of the trapping potentials and the resulting disturbance is observed for some finite length of time. Thus the quantity we are actually interested in experimentally is not $\delta\Phi\rt$, but rather something like $\Theta(0) \delta\Phi\rt \Theta(T_{\text{obs}}-t)$, where $T_{\text{obs}}$ is the experimental observation time and $\Theta(t)$ is the unit step function. Modelling $\Theta(T_{\text{obs}}-t)$ by the decaying exponential $e^{-t/T_{\text{obs}}}$ we therefore want to calculate
\beq
\overline{\delta\Phi}\rt = \delta\Phi\rt e^{-\gamma t}\Theta(0),
\eeq
with $\gamma = 1/T_{\text{obs}}$ and $\delta\Phi(\bfr,0)=0$. The Fourier transform of this quantity amounts to putting $\omega \rightarrow \omega + i \gamma$ in the coefficients $\{b_i(\omega)\}$. This replacement can be neglected in the perturbation $P\rw$ because this is not divergent and will generally have an intrinsic width greater than $\gamma$ in any case.

If more accurate modelling of the experimental resolution is required, numerical results for the condensate response in the frequency domain can be transformed to the time-domain at the end of the calculation. The factor $e^{-\gamma t}$ can then be removed and some other weight function introduced as appropriate. Of course, when calculating decay rates by fitting Lorentzians to the calculated spectra, the explicit contribution from $\gamma$ in the denominator of $\mathcal{G}_p(\omega+i\gamma)$ in Eq.~(\ref{G}) must be subtracted. The remaining dependence of the decay rate and energy shift on $\gamma$ comes from the self-energy $\Sigma_p(\omega+i\gamma)$ of Eq.~(\ref{sigmapw}). In practice this dependence is usually weak and numerical results do not depend significantly on $\gamma$ for experimentally relevant values (typically of order a few times $10^{-2}$ of the trapping frequency).

Of course, there may be sources of broadening other than experimental resolution which can conveniently be included in $\gamma$, such as collisions with background gas atoms for example. In principle, we could also model non-condensate collisions as an additional source of broadening of the quasiparticle levels. This would only affect the value of $\gamma$ in the dynamic terms $\Delta E_p^{(D)}(\omega+i\gamma)$ and $\Delta P_{p0}^{(D)}(\omega+i\gamma)$ and would involve having two values of $\gamma$ in the theory, an explicit contribution $\gamma_1$ in the denominator of $\mathcal{G}_p$ and a larger value $\gamma_2$ in the dynamic terms. If this had a significant effect, however, such processes would have to be described more rigorously via a higher order extension of the theory.

\section{Variance in the condensate population} \label{app_condensate_statistical_fluctuations}

To calculate $\Delta N_0$ of Eq.~(\ref{DN_0}), we need an expression for the variance in the condensate population $\text{Var}(N_0) = \langle \hat{N}_0^2 \rangle - \langle \hat{N}_0 \rangle^2$, where the averages are calculated in the quasiparticle basis. For a system with a fixed total number of atoms we have $\text{Var}(N_0) = \text{Var}(N_{\text{nc}}) = \langle \hat{N}_{\text{nc}}^2 \rangle - \langle \hat{N}_{\text{nc}} \rangle^2$, where $\hat{N}_{\text{nc}} = \intdr \hat{\Lambda}^{\dagger}\rt\hat{\Lambda}\rt$ is the non-condensate number operator. Dropping the time label, $\langle \hat{N}_{\text{nc}}^2 \rangle$ is therefore defined by
\beq
\langle \hat{N}_{\text{nc}}^2 \rangle= \iint \! \! d^3\bfr \, d^3\bfrp \left\langle \hat{\Lambda}^{\dagger}(\bfr)\hat{\Lambda}(\bfr)\hat{\Lambda}^{\dagger}(\bfrp)\hat{\Lambda}(\bfrp) \right\rangle.
\eeq

This can be calculated in equilibrium using Wick's theorem which gives
\begin{align}
\left\langle \hat{\Lambda}^{\dagger}(\bfr)\hat{\Lambda}(\bfr)\hat{\Lambda}^{\dagger}(\bfrp)\hat{\Lambda}(\bfrp)\right\rangle &= \nt(\bfr)\nt(\bfrp) + |\mt(\bfr,\bfrp)|^2 \\
&+ \nt(\bfrp,\bfr)\left [\nt(\bfr,\bfrp) + Q(\bfr,\bfrp) \right].\nonumber
\end{align}
We therefore have
\beq
\text{Var}(N_{\text{nc}}) = N_{\text{nc}} + \iint \! \! d^3\bfr \, d^3\bfrp \,\, |\nt(\bfr,\bfrp)|^2 + |\mt(\bfr,\bfrp)|^2,
\label{app_dN02}
\eeq
where $N_{\text{nc}} = \langle \hat{N}_{\text{nc}} \rangle = \intdr \nt(\bfr)$. After some algebra this reduces to
\begin{align}
\text{Var}(N_{\text{nc}}) &= \sum_{ij>0} \left [ N_iN_j+ \frac{(N_i+N_j)}{2} \right ]\left [ |I_{ij}|^2 + |J_{ij}|^2 \right ] \nonumber \\
&+ \sum_{ij>0} 2 U_{ij}V_{ij}^*,
\end{align}
where $I_{ij}$, $J_{ij}$, are defined in Eqs.~(\ref{Iij}) and (\ref{Jij}) while $U_{ij}$ and $V_{ij}$ are defined in Eqs.~(\ref{Uij}) and (\ref{Vij}). In the homogeneous limit this expression agrees with the result obtained in \cite{Giorgini98a}.


\begin{thebibliography}{99}

\bibitem{Jin96}  D. S. Jin, J. R. Ensher, M. R. Matthews, C. E. Wieman, and E. A. Cornell, Phys. Rev. Lett. {\bf 77}, 420 (1996).

\bibitem{Jin97}D. S. Jin, M. R. Matthews, J. R. Ensher, C. E. Wieman, and E. A. Cornell, Phys. Rev. Lett. {\bf 78}, 764 (1997).

\bibitem{Mewes96b}M.-O. Mewes, M. R. Andrews, N. J. van
Druten, D. M. Kurn, D. S. Durfee, C. G. Townsend, and W. Ketterle, Phys. Rev. Lett. {\bf 77}, 988 (1996).

\bibitem{StamperKurn98}D. M. Stamper-Kurn, H.-J. Miesner, S. Inouye, M. R. Andrews, and W. Ketterle, Phys. Rev. Lett. {\bf 81}, 500 (1998).

\bibitem{Marago01}O. Marag\`{o}, G. Hechenblaikner, E. Hodby, and C. Foot, Phys. Rev. Lett. {\bf 86}, 3938 (2001).

\bibitem{Chevy02}F. Chevy, V. Bretin, P. Rosenbusch, K. W. Madison, and J. Dalibard, Phys. Rev. Lett. {\bf 88}, 250402 (2002).

\bibitem{Steinhauer03}J. Steinhauer, N. Katz, R. Ozeri, N. Davidson, C. Tozzo, and F. Dalfovo, Phys. Rev. Lett. {\bf 90}, 060404 (2003).

\bibitem{Bretin03}V. Bretin, P. Rosenbusch, F. Chevy, G. V. Shlyapnikov, and J. Dalibard, Phys. Rev. Lett. {\bf 90}, 100403 (2003).

\bibitem{Morgan03a}S. A. Morgan, M. Rusch, D. A. W. Hutchinson, and K. Burnett, cond-mat/0305535.

\bibitem{Bogoliubov47}N. N. Bogoliubov, J. Phys. U.S.S.R. {\bf 11}, 23 (1947).

\bibitem{Hugenholtz59}N. M. Hugenholtz and D. Pines, Phys. Rev. {\bf 116}, 489 (1959).

\bibitem{GPE}E. P. Gross, Nuovo Cimento {\bf 20}, 454 (1961); J. Math. Phys. {\bf 4}, 195 (1963). L. P. Pitaevskii, Sov. Phys. JETP {\bf 13}, 451 (1961).

\bibitem{Reidl00}J. Reidl, A. Csord\'{a}s, R. Graham, and P. Sz\'{e}pfalusy, Phys. Rev. A {\bf 61}, 043606 (2000).

\bibitem{Jackson02}B. Jackson and E. Zaremba, Phys. Rev. Lett. {\bf 88}, 180402 (2002).

\bibitem{Morgan03c}S. A. Morgan, (unpublished).

\bibitem{Beliaev58}S. T. Beliaev, Sov. Phys. JETP {\bf 7}, 289, (1958); {\bf 7}, 299, (1958).

\bibitem{Popov65a}V. N. Popov and L. D. Fadeev, Sov. Phys. JETP {\bf 20}, 890 (1965).

\bibitem{Popov65b}V. N. Popov, Sov. Phys. JETP {\bf 20}, 1185 (1965).

\bibitem{deGennes66}P. R. de Gennes, {\em Superconductivity of Metals and Alloys} (Benjamin, New York, 1966).

\bibitem{Fetter72} A. L. Fetter, Ann. Phys. (N.Y.) {\bf 70}, 67 (1972).

\bibitem{Shi98} H. Shi and A. Griffin, Phys. Rep. {\bf 304}, 1 (1998).

\bibitem{Fedichev98}P. O. Fedichev and G. V. Shlyapnikov, Phys. Rev. A {\bf 58}, 3146 (1998).

\bibitem{Mohling60}F. Mohling and A. Sirlin, Phys. Rev. {\bf 118}, 370 (1960); F. Mohling and M. Morita, \textit{ibid.} {\bf 120}, 681 (1960).

\bibitem{Morgan00}S. A. Morgan, J. Phys. B {\bf 33}, 3847 (2000).

\bibitem{Ruprecht96}P. A. Ruprecht, M. Edwards, K. Burnett, and C. W. Clark, Phys. Rev. A {\bf 54}, 4178 (1996).

\bibitem{Edwards96} M. Edwards, P. A. Ruprecht, K. Burnett, R. J. Dodd, and C. W. Clark, Phys. Rev. Lett. {\bf 77}, 1671 (1996).

\bibitem{Stringari96}S. Stringari, Phys. Rev. Lett. {\bf 77}, 2360 (1996).

\bibitem{Giorgini00}S. Giorgini, Phys. Rev. A {\bf 61}, 063615 (2000).

\bibitem{Minguzzi97}A. Minguzzi and M. P. Tosi, J. Phys.: Condens. Matter {\bf 9}, 10211 (1997).

\bibitem{Guilleumas99}M. Guilleumas and L. P. Pitaevskii, Phys. Rev. A {\bf 61}, 013602 (1999).

\bibitem{Rusch99}M. Rusch and K. Burnett, Phys. Rev. A {\bf 59}, 3851 (1999).

\bibitem{Bene98}Gy. Bene and P. Sz\'{e}pfalusy, Phys. Rev. A {\bf 58}, R3391 (1998).

\bibitem{Davis01}M. J. Davis, S. A. Morgan, and K. Burnett, Phys. Rev. Lett. {\bf 87}, 160402 (2001).

\bibitem{Davis01b}M. J. Davis, R. J. Ballagh, and K. Burnett, J. Phys. B {\bf 34}, 4487 (2001).

\bibitem{Stoof01}H. T. C. Stoof and M. J. Bijlsma, J. Low. Temp. Phys. {\bf 124}, 431 (2001).

\bibitem{Duine01}R. A. Duine and H. T. C. Stoof, Phys. Rev. A {\bf 65}, 013603 (2001).

\bibitem{Gardiner02}C. W. Gardiner, J. R. Anglin, and T. I. A. Fudge, J. Phys. B {\bf 35}, 1555 (2002).

\bibitem{Griffin96}A. Griffin, Phys. Rev. B {\bf 53}, 9341 (1996).

\bibitem{Hutchinson97}D. A. W. Hutchinson, E. Zaremba, and A. Griffin, Phys. Rev. Lett {\bf 78}, 1842 (1997).

\bibitem{Dodd98} R. J. Dodd, M. Edwards, C. W. Clark, and K. Burnett, Phys. Rev. A {\bf 57}, R32 (1998).

\bibitem{Reidl99}J. Reidl, A. Csord\'{a}s, R. Graham, and P. Sz\'{e}pfalusy, Phys. Rev. A {\bf 59}, 3816 (1999).

\bibitem{Proukakis98}N. P. Proukakis, S. A. Morgan, S. Choi, and K. Burnett, Phys. Rev. A {\bf 58}, 2435 (1998).

\bibitem{Hutchinson98}D. A. W. Hutchinson, R. J. Dodd, and K. Burnett, Phys. Rev. Lett {\bf 81}, 2198 (1998).

\bibitem{Hutchinson00}D. A. W. Hutchinson, K. Burnett, R. J. Dodd, S. A. Morgan, M. Rusch, E. Zaremba, N. P. Proukakis, M. Edwards, and C. W. Clark, J. Phys B \textbf{33}, 3825 (2000).

\bibitem{Bijlsma99}M. J. Bijlsma and H. T. C. Stoof, Phys. Rev. A {\bf 60}, 3973 (1999).

\bibitem{Khawaja00}U. A. Khawaja and H. T. C. Stoof, Phys. Rev. A {\bf 62}, 053602 (2000).

\bibitem{Jackson01}B. Jackson and E. Zaremba, Phys. Rev. Lett. {\bf 87}, 100404 (2001).

\bibitem{Zaremba99}E. Zaremba, T. Nikuni, and A. Griffin, J. Low Temp. Phys. {\bf 116}, 277 (1999).

\bibitem{Holland01}M. Holland, J. Park, and R. Walser, Phys. Rev. Lett. {\bf 86}, 1915 (2001).

\bibitem{Kohler03}T. K\"{o}hler, T. Gasenzer, and K. Burnett, Phys. Rev. A {\bf 67}, 013601 (2003).

\bibitem{Hodby01} E. Hodby, O. M. Marag\`{o}, G. Hechenblaikner, and C. J. Foot, Phys. Rev. Lett. {\bf 86}, 2196 (2001).

\bibitem{Katz02} N. Katz, J. Steinhauer, R. Ozeri, and N. Davidson, Phys. Rev. Lett. {\bf 89}, 220401 (2002).

\bibitem{Mizushima03}T. Mizushima, M. Ichioka, and K. Machida, Phys. Rev. Lett. {\bf 90}, 180401 (2003).

\bibitem{Walser99}R. Walser, J. Williams, J. Cooper, and M. Holland, Phys. Rev. A {\bf 59}, 3878 (1999).

\bibitem{Walser00}R. Walser, J. Cooper, and M. Holland, Phys. Rev. A {\bf 63}, 013607 (2000).

\bibitem{Wachter02}J. Wachter, R. Walser, J. Cooper, and M. Holland, cond-mat/0212432.

\bibitem{Giorgini98}S. Giorgini, Phys. Rev. A {\bf 57}, 2949 (1998).

\bibitem{Williams01}J. E. Williams and A. Griffin, Phys. Rev. A, {\bf 63}, 023612 (2001); \textit{ibid.} {\bf 64}, 013606 (2001).

\bibitem{Fedichev98b}P. O. Fedichev, G. V. Shlyapnikov, and J. T. M. Walraven, Phys. Rev. Lett. {\bf 80}, 2269 (1998).

\bibitem{Pitaevskii97} L. P. Pitaevskii and S. Stringari, Phys. Lett. A {\bf 235}, 398 (1997).

\bibitem{Castin98}Y. Castin and R. Dum, Phys. Rev. A {\bf 57}, 3008 (1998).

\bibitem{Rusch00}M. Rusch, S. A. Morgan, D. A. W. Hutchinson, and K. Burnett, Phys. Rev. Lett. {\bf 85}, 4844 (2000).

\bibitem{Girardeau59} M. Girardeau and R. Arnowitt, Phys. Rev. {\bf 113}, 755 (1959). See also M. Girardeau, Phys. Rev. A {\bf 58}, 775 (1998).

\bibitem{Gardiner97} C. W. Gardiner, Phys. Rev. A {\bf 56}, 1414 (1997).

\bibitem{Morgan03d}S. A. Morgan and S. A. Gardiner, (unpublished).

\bibitem{Huang_StatMech} K. Huang, {\em Statistical Mechanics}, 2nd ed. (Wiley, New York, 1987).

\bibitem{Stoof93}H. T. C. Stoof and M. Bijlsma, Phys. Rev. E {\bf 47}, 939 (1993).

\bibitem{Olshanii02} M. Olshanii and L. Pricoupenko, Phys. Rev. Lett. {\bf 88}, 010402 (2002).

\bibitem{Penrose56}O. Penrose and L. Onsager, Phys. Rev. {\bf 104}, 576 (1956).

\bibitem{Blaizot_Ripka} J.-P. Blaizot and G. Ripka, {\em Quantum Theory of Finite Systems} (MIT Press, Cambridge MA, 1986).

\bibitem{Dodd97}R. J. Dodd, K. Burnett, M. Edwards, and C. W. Clark, Phys. Rev. A {\bf 56}, 587 (1997).

\bibitem{Isoshima99}T. Isoshima and K. Machida, Phys. Rev. A {\bf 59}, 2203 (1999).

\bibitem{Svidzinksy00} A. A. Svidzinksy and A. L. Fetter, Phys. Rev. Lett. {\bf 84}, 5919 (2000).

\bibitem{Feder01} D. L. Feder, A. A. Svidzinksy, A. L. Fetter, and C. W. Clark, Phys. Rev. Lett. {\bf 86}, 564 (2001).

\bibitem{Virtanen01}S. M. M. Virtanen, T. P. Simula, and M. M. Salomaa, Phys. Rev. Lett. {\bf 86}, 2704 (2001).

\bibitem{Proukakis98b}N. P. Proukakis, K. Burnett, and H. T. C. Stoof, Phys. Rev. A {\bf 57}, 1230 (1998).

\bibitem{Kohler02}T. K\"{o}hler and K. Burnett, Phys. Rev. A {\bf 65}, 033601 (2002).

\bibitem{Bijlsma97}M. Bijlsma and H. T. C. Stoof, Phys. Rev. A {\bf 55}, 498 (1997).

\bibitem{gapless_coefficients} These coefficients are related to those defined in Eqs.~(75) and (76) of \cite{Morgan00} by (a tilde denotes coefficients from \cite{Morgan00} obtained using the contact potential for binary interactions): $A_{pij}$ =  $\tilde{A}_{pij} + \text{(all permutations of $pij$)}$, $B1_{pij}$ =  $\tilde{B}_{pij}^* + \tilde{B}_{pji}^*$ and $B2_{pij}$ =  $\tilde{B}_{ijp} + \tilde{B}_{ipj}$.

\bibitem{Giorgini98a}S. Giorgini, L. P. Pitaevskii, and S. Stringari, Phys. Rev. Lett. {\bf 80}, 5040 (1998).

\bibitem{Dobson94} J. F. Dobson, Phys. Rev. Lett. {\bf 73}, 2244 (1994).

\bibitem{Morgan97}S. A. Morgan, R. J. Ballagh and K. Burnett, Phys. Rev. A {\bf 55}, 4338 (1997).

\bibitem{Fetter98}A. L. Fetter and D. Rokhsar, Phys. Rev. A {\bf 57}, 1191 (1998).

\end{thebibliography}

\end{document}